\documentclass[12pt]{article}
\usepackage{amsmath,amssymb,amsthm,amsxtra,overpic,bbm,bm,epsfig,subfigure}
\usepackage{color}
\usepackage{comment}
\usepackage[all,cmtip]{xy}
\usepackage{cite}
\usepackage{array}
\usepackage{caption}
\usepackage{multirow}
\usepackage{slashed,stmaryrd}

\begin{document}

\vspace{0.2cm}

\begin{center}
{\large\bf Production and constraints for a massive dark photon at electron-positron colliders}
\end{center}

\vspace{0.2cm}

\begin{center}

{\bf Jun Jiang}~$^a$ \footnote{E-mail: jiangjun87@sdu.edu.cn}, \quad {\bf Chun-Yuan Li}~$^a$ \footnote{E-mail: lichunyuan@mail.sdu.edu.cn}, \quad {\bf Shi-Yuan Li}~$^a$ \footnote{E-mail: lishy@sdu.edu.cn}, \quad {\bf Shankar Dayal Pathak}~$^{a, b}$ \footnote{E-mail: prince.pathak19@gmail.com}, \quad {\bf Zong-Guo Si}~$^a$ \footnote{E-mail: zgsi@sdu.edu.cn}, \quad {\bf Xing-Hua Yang}~$^a$  \footnote{E-mail: yangxh@mail.sdu.edu.cn}

{$^a$School of Physics, Shandong University, Jinan, Shandong 250100, China}\\
{$^b$ Department of Physics,
Lovely Professional University,
Phagwara, Punjab, 144411, India}

\end{center}

\vspace{0.2cm}
\begin{abstract}

Dark sector may couple to the Standard Model via one or more mediator particles. We discuss two types of mediators: the dark photon $A^{\prime}$ and the dark scalar mediator $\phi$. The total cross-sections and various differential distributions of the processes $e^{+} e^{-} \rightarrow q \bar{q} A^{\prime}$ and $e^{+} e^{-} \rightarrow q \bar{q} \phi$ ($q=u,~d,~c,~s$ and $b$ quarks) are discussed. We focus on the study of the invisible $A^{\prime}$ due to the cleaner background at future $e^{+} e^{-}$ colliders. It is found that the kinematic distributions of the two-jet system could be used to identify (or exclude) the dark photon and the dark scalar mediator, as well as to distinguish between them. We further study the possibility of a search for dark photons at a future CEPC experiment with $\sqrt{s}=$ 91.2~GeV and 240~GeV. With CEPC running at $\sqrt{s}=$ 91.2~GeV, it would be possible to perform a decisive measurement of the dark photon (20~GeV $< m_{A^{\prime}} <$ 60~GeV) in less than one operating year. The lower limits of the integrated luminosity for the significance $S/\sqrt{B}= 2\sigma,~3\sigma~\textrm{and}~5\sigma$  are presented.

\end{abstract}

\section{Introduction}

Signals of non-baryonic dark matter (DM) in the Universe have been identified in a number of astrophysical and cosmological observations, such as the Cosmic Microwave Background anisotropy measurements, galactic rotation curves, large scale structure surveys, X-ray observations and gravitational lensing \cite{Aghanim:2018eyx,Conroy:2010bs,Sofue:2000jx,Cole:2005sx,Beutler:2011hx,Anderson:2012sa,Vikhlinin:2008ym,Fu:2007qq,Massey:2007gh,Jungman:1995df,Bertone:2004pz}.
The contribution of DM is nearly  $75\%$ of the total matter in the Universe.
Specifically, the Planck data give the value of the relic density of DM of $\Omega_{CDM}h^{2} = 0.120 \pm 0.001$~\cite{Aghanim:2018eyx}.
 DM influences the dynamical effects from the scale of a galaxy up to the cosmic scale, and plays a crucial role in the galaxy rotation curve and the formation of structures in the Universe.
However, the nature of the DM particles remains a mystery and has become one of the most important challenges of modern science. The underlying physics of DM particles is explored by various worldwide projects, such as the direct and indirect searches, collider experiments and astrophysical signatures arising from DM self-interactions ~\cite{Undagoitia:2015gya,Gaskins:2016cha,Kahlhoefer:2017dnp,Tulin:2017ara}.

Given the intricate structure of the Standard Model (SM), which describes only a sub-dominant component of the Universe, it would not be surprising if the dark sector contains itself a rich structure, with DM making only a part of it. In the dark sector, the DM particles do not interact directly with the known forces, except with the gravitational force. However, there are typically one or more mediator particles which are coupled with SM and act as a ``portal"~\cite{Chu:2011be,Alexander:2016aln,Dutra:2018gmv,Essig:2013lka,Evans:2017kti,Pospelov:2007mp}.
Such extended interactions associating the dark sector and SM depend on the spin and parity: the mediators can be vector $A^{\prime}$, scalar $\phi$, pseudoscalar $a$, axial-vector $Z^{\prime}$ and even fermions $N$.

A new force mediated by dark photons has been a subject of considerable interest in high energy physics. The existence of the dark photon ~\cite{Pospelov:2008zw,Holdom:1985ag,Cirelli:2016rnw}, associated to a hidden $U(1)^{\prime}$ gauge interaction, has been the subject of many investigations, both theoretical and experimental. Substantial effort has been invested in the search for dark photons using various processes including bremsstrahlung $e^{-}\textrm{Z} \rightarrow e^{-}\textrm{Z}A^{\prime}$~\cite{Banerjee:2017hhz,Abrahamyan:2011gv,Merkel:2014avp,Gninenko:2017yus},
 meson decays $\pi^{0}/\eta/\eta^{\prime} \rightarrow \gamma A^{\prime}$, $K \rightarrow \pi A^{\prime}$, $\phi \rightarrow \eta A^{\prime}$ and $D^{\ast} \rightarrow D^{0} A^{\prime}$
~\cite{Adare:2014mgk,Feng:2017uoz,Ilten:2015hya}, the Drell-Yan process $q \bar q \rightarrow A^{\prime} \rightarrow (\ell^{+} \ell^{-} ~\textrm{or} ~h^{+} h^{-})$~\cite{Curtin:2014cca, Chatrchyan:2013tia}, annihilation $e^{+} e^{-} \rightarrow \gamma A^{\prime}$~\cite{Lees:2017lec,Fayet:2007ua,He:2017zzr,Jiang:2018jqp,Anastasi:2016ktq}, etc. Stringent limits for the kinetic mixing parameter $\varepsilon$ for a given dark photon mass $m_{A^{\prime}}$ have been obtained ~\cite{Essig:2013lka,Ilten:2016tkc,Cirelli:2016rnw,Evans:2017kti,Prasad:2019ris}.
For $m_{A^{\prime}}\lesssim 1$~GeV, only limited values of $\varepsilon$ are allowed. However, for a heavy dark photon, a wide range of mixing parameter values has not been excluded by the current experiments.

Future high-energy electron-positron colliders provide an opportunity to search for the dark sector mediators.
These colliders include CEPC ~\cite{CEPCStudyGroup:2018ghi}, ILC ~\cite{Baer:2013cma}, FCC-ee ~\cite{Gomez-Ceballos:2013zzn} and CLIC ~\cite{Abramowicz:2013tzc}, with the center-of-mass energy $\sqrt{s}$ varying from 91.2 GeV to 1 TeV.
Assuming that dark mediators interact only with quarks, we investigate in this work the production of dark photon $A^{\prime}$ and dark scalar mediator $\phi$ at electron-positron colliders with $\sqrt{s} =$ 91.2 GeV, 240 GeV, 500 GeV and 1 TeV. We analyze the cross-sections and the normalized kinematic distributions of the processes $e^{+}e^{-}\rightarrow q \bar{q} A^{\prime}$ and $e^{+}e^{-}\rightarrow q \bar{q} \phi$, and focus on the invisible $A^{\prime}$ due to a cleaner background. The corresponding background processes are also simulated.

The paper is organized as follows. In Sec. ~2, we present a simple theoretical framework for the dark photon and dark scalar mediator. In Sec.~3, we investigate the production of dark photon and dark scalar mediator at future $e^{+}e^{-}$ colliders, and discuss how to distinguish between them. In Sec.~4, we study the discovery potential
of dark photons at a CEPC experiment. Finally, a short summary is given.

\section{Dark photon and Dark scalar mediator}

In a simple extension of SM, one can introduce a $U(1)^{\prime}$ as an extra gauge group. The gauge boson $A^{\prime}$ arises from the extra $U(1)^{\prime}$ gauge group, which can be coupled weakly to electrically charged particles by ``kinetic mixing" with the photon~\cite{Pospelov:2008zw,Holdom:1985ag,Cirelli:2016rnw}. Kinetic mixing produces an effective parity-conserving interaction $\varepsilon e A^{\prime}_{\mu}J^{\mu}_{EM}$ of $A^{\prime}$ with the electromagnetic current $J^{\mu}_{EM}$, suppressed relative to the electron charge by the parameter $\varepsilon$~\cite{Essig:2013lka}.
The gauge boson or dark photon $A'$ play the role of the ``vector portal" connecting the SM and DM particles. We assume that the dark photon only interacts with the DM particles and SM quarks. After diagonalization of the kinetic mixing term, the Lagrangian of the dark photon model is ~\cite{Dutra:2018gmv,Alexander:2016aln}
\begin{eqnarray}
{\cal L}&\supset& \sum_{q}\bar{q}(-ec_{q}\gamma^{\mu}A_{\mu}-\varepsilon ec_{q}\gamma^{\mu}A^{\prime}_{\mu}-m_{q})q+\bar{\chi}(-g_{\chi}\gamma^{\mu}A^{\prime}_{\mu}-m_{\chi})\chi \nonumber\\
&&-\frac{1}{4}F_{\mu\nu}F^{\mu\nu}-\frac{1}{4}F^{\prime}_{\mu\nu}F^{\prime\mu\nu}+\frac{1}{2}m_{A'}^{2}A'^{2},\label{Lagrangian_v}
\end{eqnarray}
where $m_{q}$, $m_{\chi}$ and $m_{A^{\prime}}$ denote the masses of SM quarks, DM particle and dark photon, respectively. $c_{q}$ is the charge of the quarks. $F^{\mu\nu}$ and $F^{\prime\mu\nu}$ are the field strengths of the ordinary photon $A$ and the dark photon $A^{\prime}$, $\varepsilon$ is the kinetic mixing parameter in the physical basis, $g_{\chi}$ is the coupling parameter between the dark photon and the dark sector, and $\alpha_{\chi}=g_{\chi}^{2}/(4\pi)$ is the dark fine structure constant.

\begin{figure}[!htbp]
\centering
\includegraphics[width=3.3in]{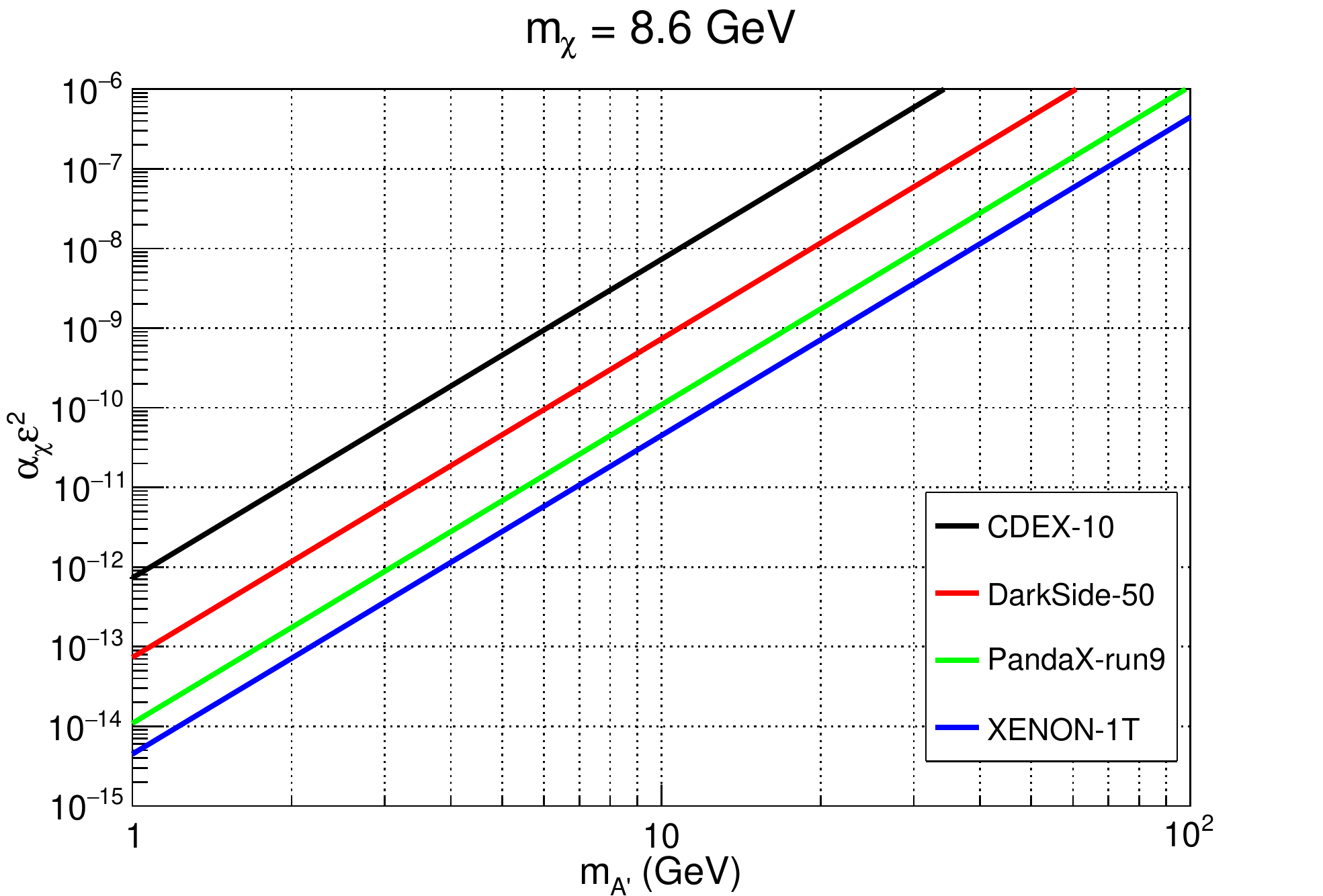}
\includegraphics[width=3.3in]{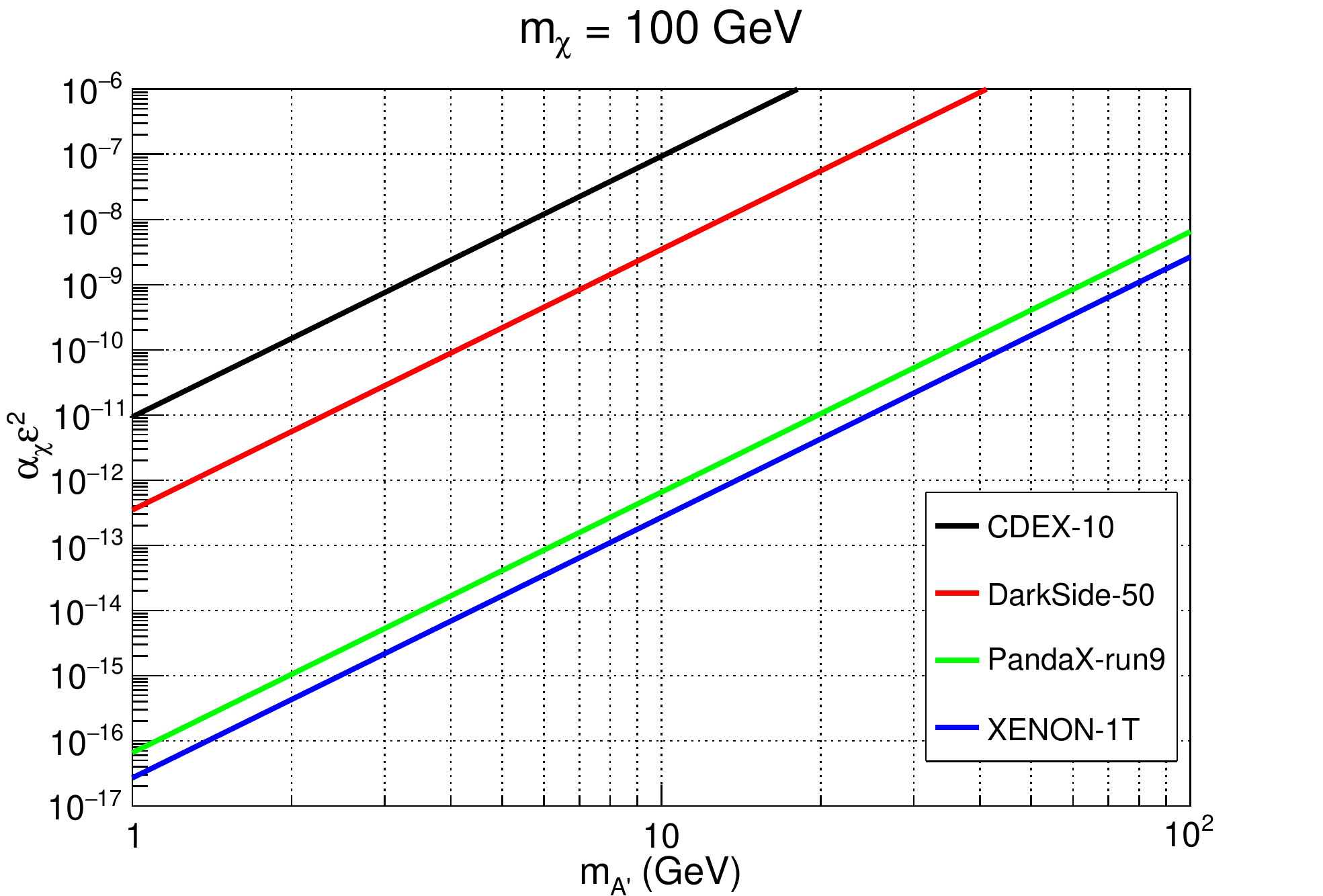}
\caption{The 90\% C.L. upper limits of the combined parameter $\alpha_{\chi}\varepsilon^{2}$ with $m_{\chi} = $ 8.6~GeV (left panel), and 100~GeV (right panel) from the CDEX-10~\cite{Jiang:2018pic}, PandaX-II~\cite{Tan:2016zwf}, DarkSide-50~\cite{Agnes:2018ves} and XENON-1T~\cite{Aprile:2018dbl} experiments.}
\label{fig:upper_limit_of_epsilon}
\end{figure}

A number of experiments have proposed restrictions on the mixing parameter $\varepsilon$ ~\cite{Essig:2013lka,Ilten:2016tkc,Cirelli:2016rnw,Evans:2017kti}. However, for the dark photon mass $m_{A^{\prime}} >$ 1~GeV, a wide range of mixing parameter values has still not been excluded by the current experiments. We can extract the maximum value of $\varepsilon$ from the direct DM detection experiments. The differential cross-sections for DM particle-nucleon scattering in the non-relativistic limit can be written as ~\cite{Cirelli:2016rnw,Fornengo:2011sz,Kaplinghat:2013yxa}
\begin{eqnarray}
 \frac{d\sigma}{dE_{R}}(v_{DM},E_{R})=\frac{8\pi\alpha_{em}\alpha_{\chi}\varepsilon^{2}m_{T}}{(2m_{T}E_{R}+m_{A^{'}}^{2})^{2}}\frac{1}{v_{DM}}Z_{T}^{2}F^{2}(2m_{T}E_{R}),
\end{eqnarray}
where $E_{R}$ is the nuclear recoil energy, $v_{DM}$ is the velocity of the DM particle in the nucleon rest frame, $\alpha_{em}=e^{2}/4\pi$ is the electromagnetic fine structure constant, $m_{T}$ is the mass of the target nucleus, $Z_{T}$ is the number of protons in the target nuclei, and $F(2m_{T}E_{R})$ is the Helm form factor~\cite{Helm:1956zz,Lewin:1995rx}. The dark fine structure constant $\alpha_{\chi}$ can be determined from the relic abundance of DM.
When $m_{\chi}$ is determined, the combined coupling parameter $\alpha_{\chi}\varepsilon^{2}$ can be constrained from the experimental data by evaluating the function $\chi^{2}= -\sum 2\rm{ln}\cal L^{\prime}$, where $\cal L^{\prime}$ is the likelihood function~\cite{Li:2014vza,Li:2019drx}.
Fig.~\ref{fig:upper_limit_of_epsilon} shows the 90\% C.L. upper limits of the combined parameter $\alpha_{\chi}\varepsilon^{2}$ with $m_{\chi} = $ 8.6~GeV (CDMS-II-Si favors a DM mass of $m_{\chi}\thicksim $ 8.6 GeV ~\cite{Agnese:2013rvf}), and $m_{\chi} = $100~GeV constrained by the CDEX-10~\cite{Jiang:2018pic}, PandaX-II~\cite{Tan:2016zwf}, DarkSide-50~\cite{Agnes:2018ves} and XENON-1T~\cite{Aprile:2018dbl} data.

 Alternatively, in the dark scalar mediator $\phi$ model, the DM particles $\chi$ can interact with the SM particles through the ``Higgs portal" ~\cite{Alexander:2016aln,Pospelov:2007mp}. The corresponding  Lagrangian can be written as,
\begin{eqnarray}
{\cal L}&\supset& \frac{1}{2}(\partial_{\mu}\phi)^{2}-\frac{1}{2}m_{\phi}^{2}\phi^{2}+\bar{\chi}(i\partial_{\mu}\gamma_{\mu}-m_{\chi}-\lambda_{\chi}\phi)\chi \nonumber\\
&&- \lambda_{1} \upsilon \phi (H^{+}H-\frac{\upsilon^{2}}{2})-\lambda_{2}\phi^{2}(H^{+}H-\frac{\upsilon^{2}}{2})-V(\phi),
\end{eqnarray}
where $H$ is the SM Higgs doublet, $\upsilon$ is the corresponding vacuum expectation value, and $\lambda_{\chi}$, $\lambda_{1}$, $\lambda_{2}$ are three parameters. In the case of $\langle\phi\rangle=0$ and $\lambda_{2}\rightarrow$ 0, after electroweak symmetry breaking, the relevant DM and mediator Lagrangian takes the following form,
\begin{eqnarray}
{\cal L}&\supset& \frac{1}{2}(\partial_{\mu}\phi)^{2}-\frac{1}{2}m_{\phi}^{2}\phi^{2}+\bar{\chi}(i\partial_{\mu}\gamma_{\mu}-m_{\chi}-\lambda_{\chi}\phi)\chi- \lambda_{1} \upsilon^{2} \phi h,
\end{eqnarray}
where the interaction between SM particles and DM particles are mediated by Higgs-singlet mixing, i.e., the $h-\phi$ scalar exchange. We assume that the dark scalar mediator $\phi$ directly couples to the SM quarks $q$. The dark scalar mediator plays a crucial role in the ``scalar portal". The mixing term can be written as $-\varepsilon_{s}e\phi q\bar {q}$. We choose $\varepsilon_{s}=\varepsilon$ for simplicity.

\section{Production of dark photon and dark scalar mediator}

\begin{figure}[!htbp]
\centering
\includegraphics[width=1.6in]{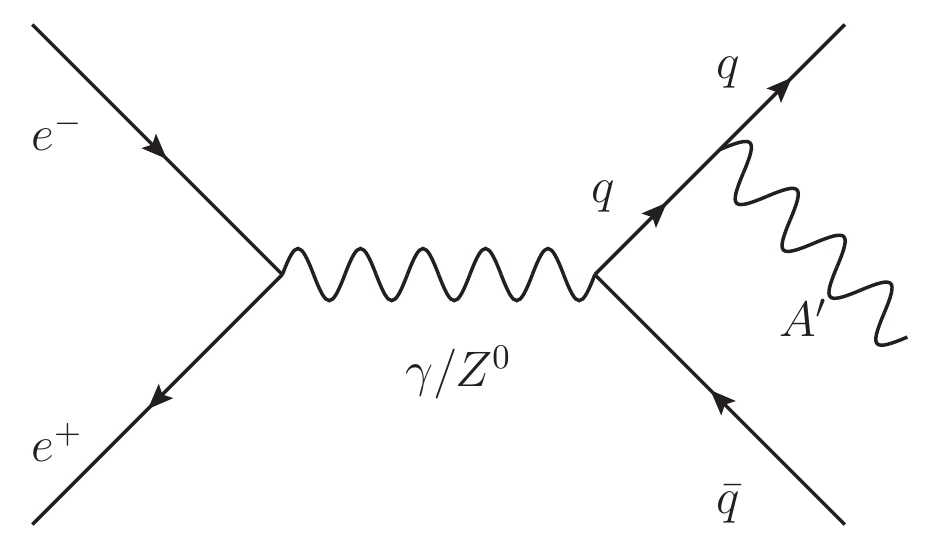}
\includegraphics[width=1.6in]{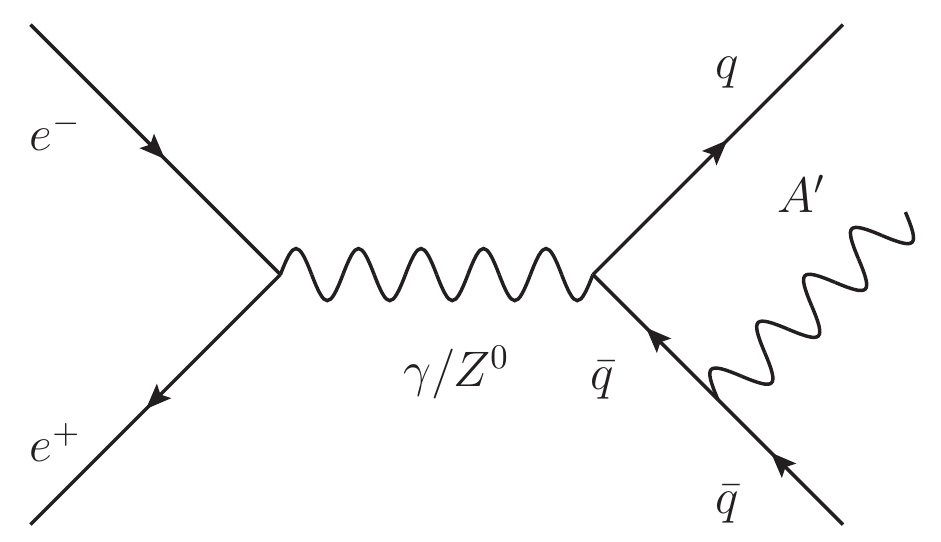}
\includegraphics[width=1.6in]{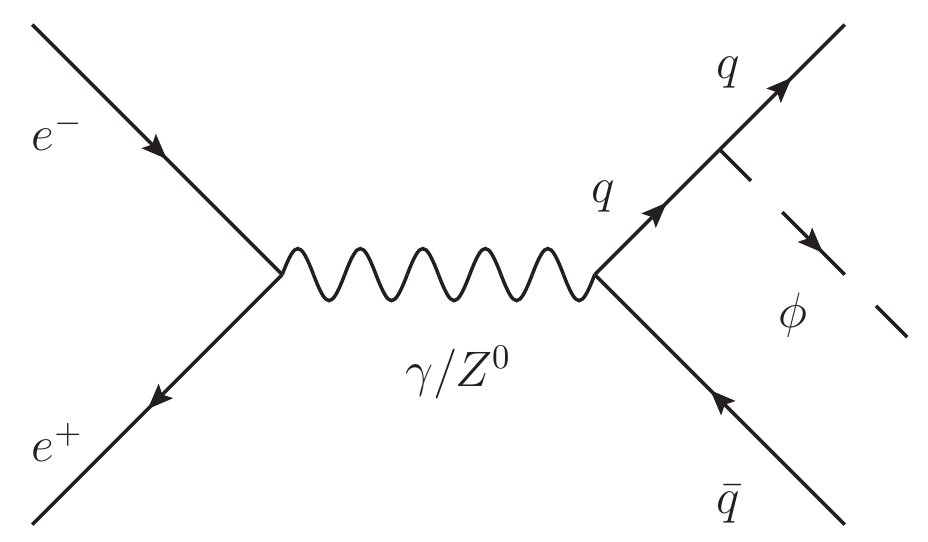}
\includegraphics[width=1.6in]{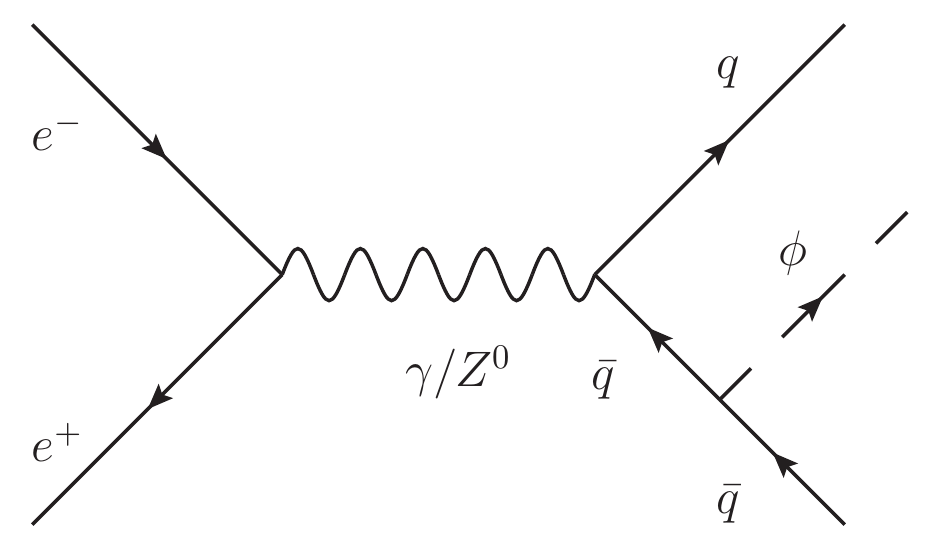}
\caption{The Feynman diagrams for the processes $e^{+} e^{-} \rightarrow q \bar{q} A^{\prime}$ and $e^{+} e^{-} \rightarrow q \bar{q} \phi$.}
\label{fig:Feynman diagrams}
\end{figure}

In this section, we investigate the production of the massive dark photon $A^{\prime}$ and of the massive dark scalar mediator $\phi$ via the processes $e^{+} e^{-} \rightarrow q \bar{q} A^{\prime}$ and $e^{+} e^{-} \rightarrow q \bar{q} \phi$ ($q=u,~d,~s,~c$ and~$b$) at the center-of-mass energies $\sqrt{s} =$ 91.2 GeV, 240 GeV, 500 GeV and 1 TeV, with different values of $m_{A^{\prime}}$ and $m_{\phi}$.
The Feynman diagrams for the production of $A'$ and $\phi$ associated with two jets at $e^+e^-$ colliders are shown in Fig.~\ref{fig:Feynman diagrams}.

\begin{figure}[!htbp]
\centering
\subfigure[]{
\includegraphics[width=3.3in]{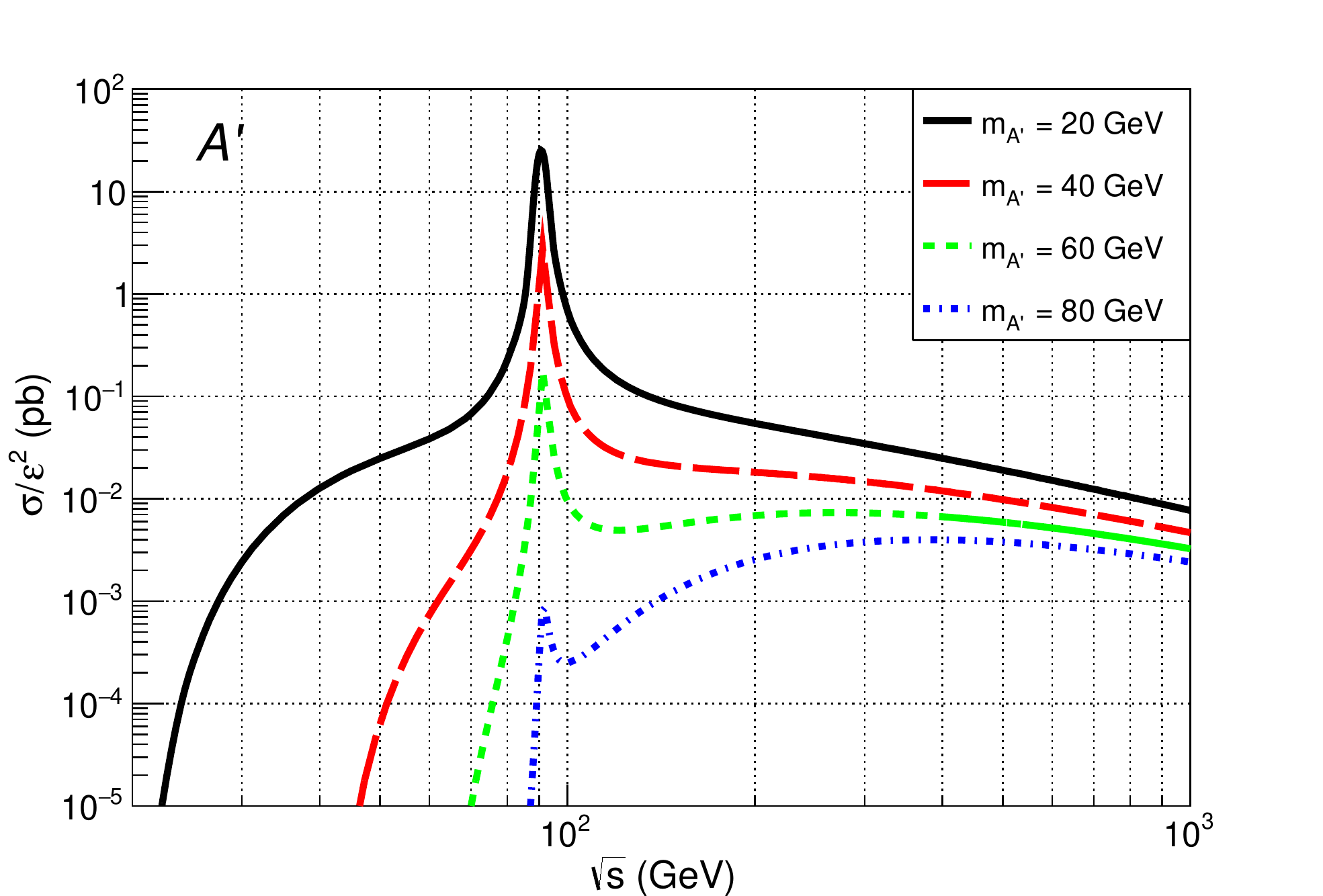}}
 \subfigure[]{
\includegraphics[width=3.3in]{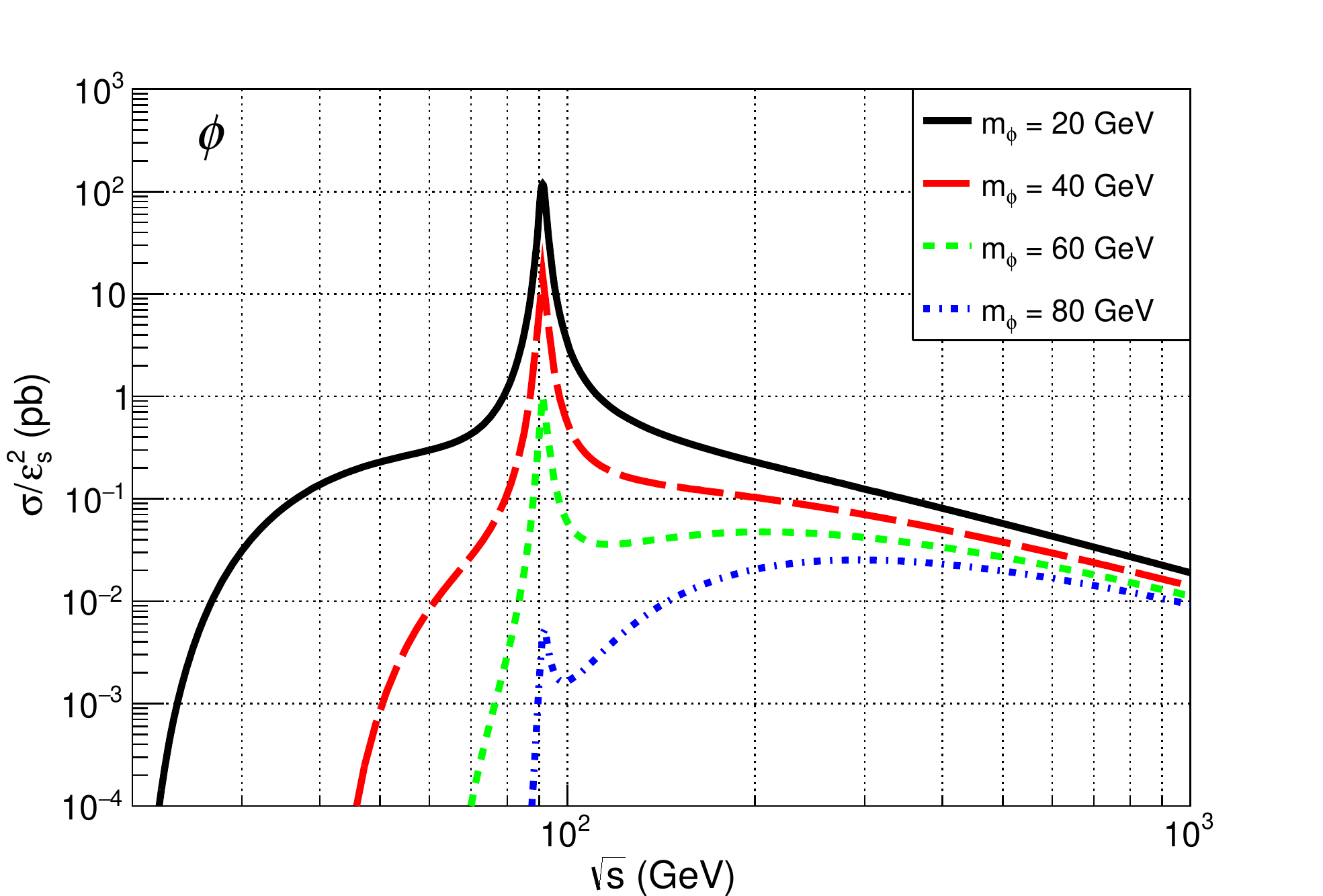}}
\subfigure[]{
\includegraphics[width=3.3in]{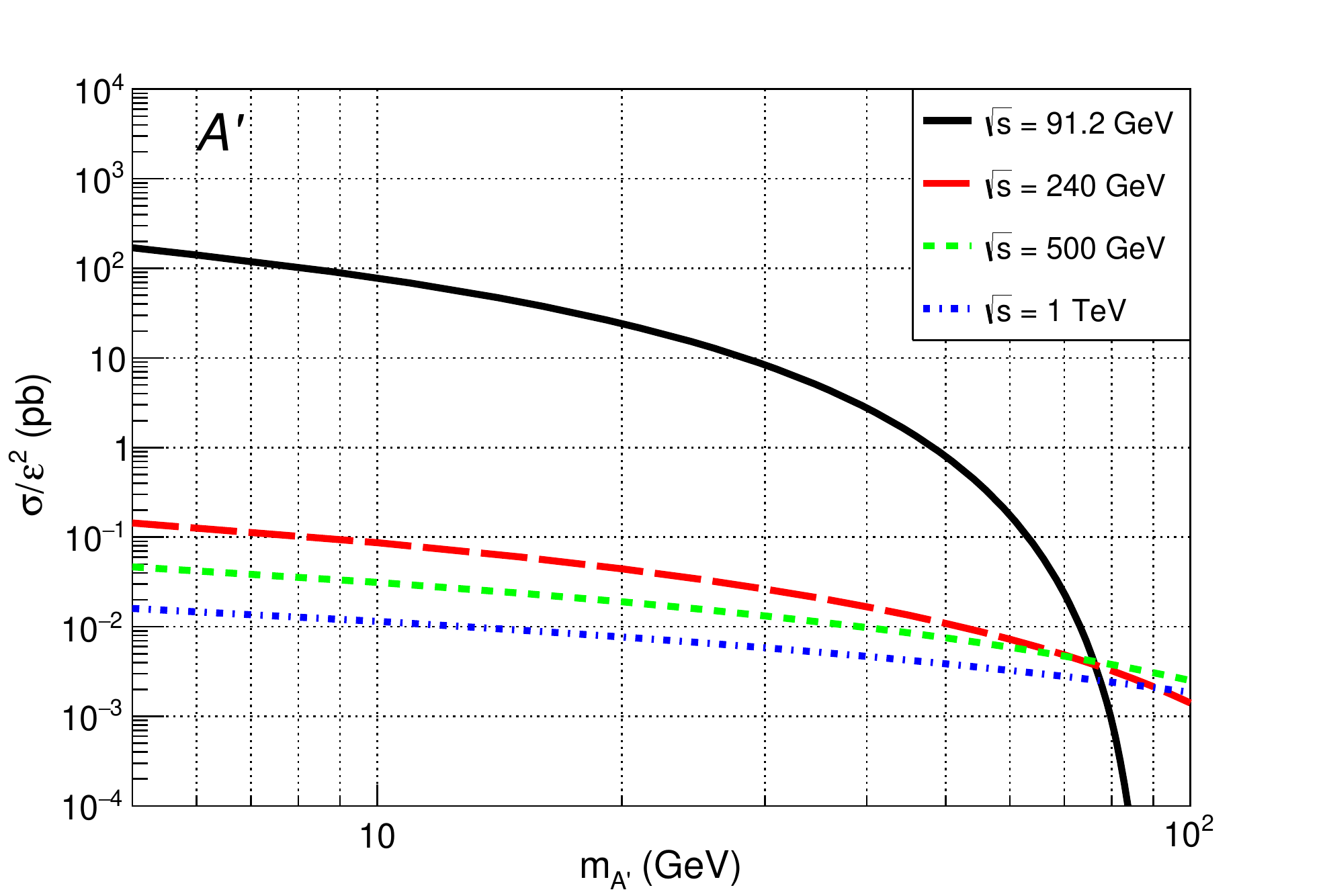}}
 \subfigure[]{
\includegraphics[width=3.3in]{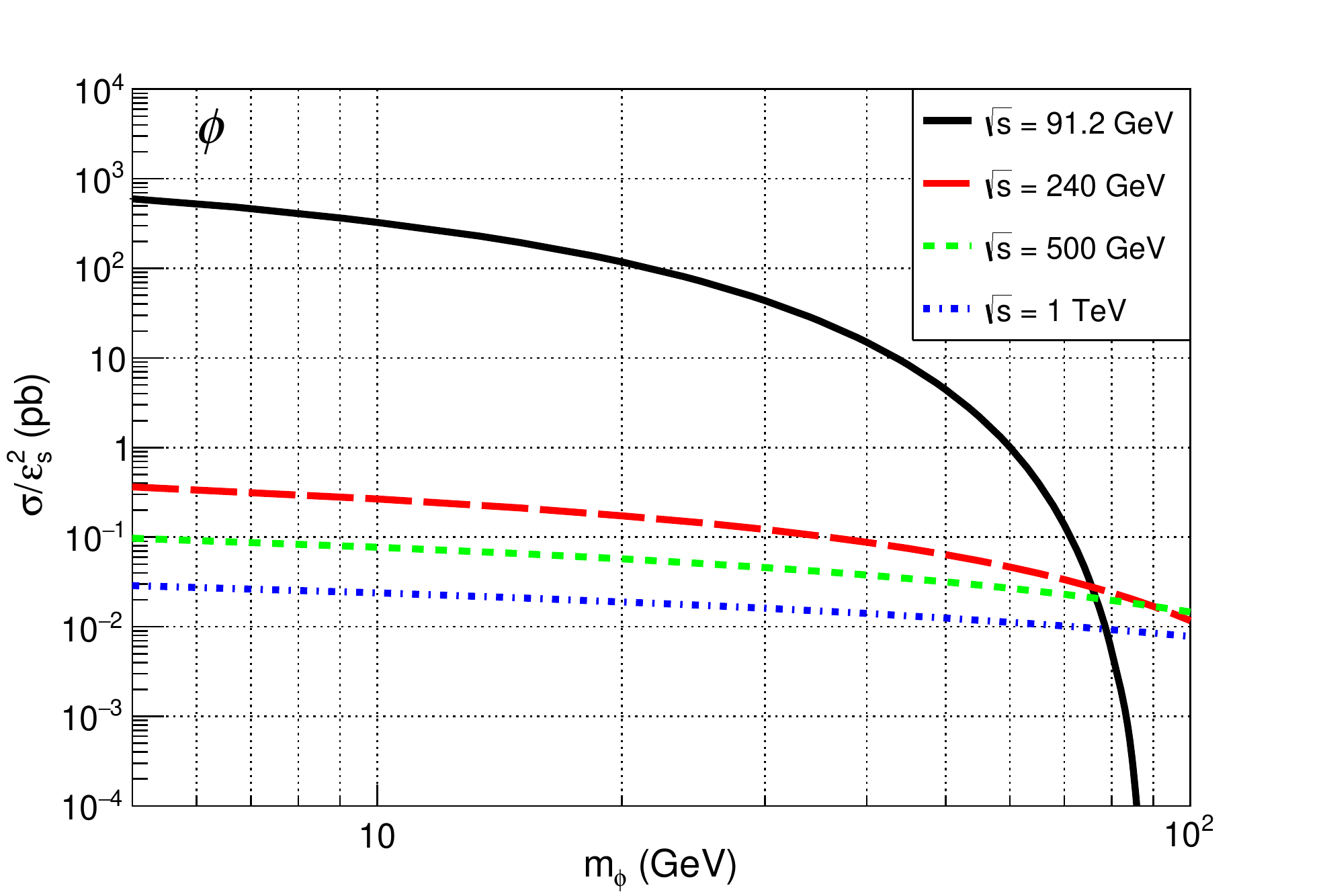}}
\caption{Reduced cross-sections of the processes $e^{+} e^{-} \rightarrow q \bar{q} A^{\prime}$ (left panels) and $e^{+} e^{-} \rightarrow q \bar{q} \phi$ (right panels) as function of $\sqrt{s}$ and $m_{A^{\prime}}$ or $m_{\phi}$.}
\label{fig:cross_section_v_s} 
\end{figure}

\begin{figure}[htbp]
\centering
\includegraphics[width=2.2in]{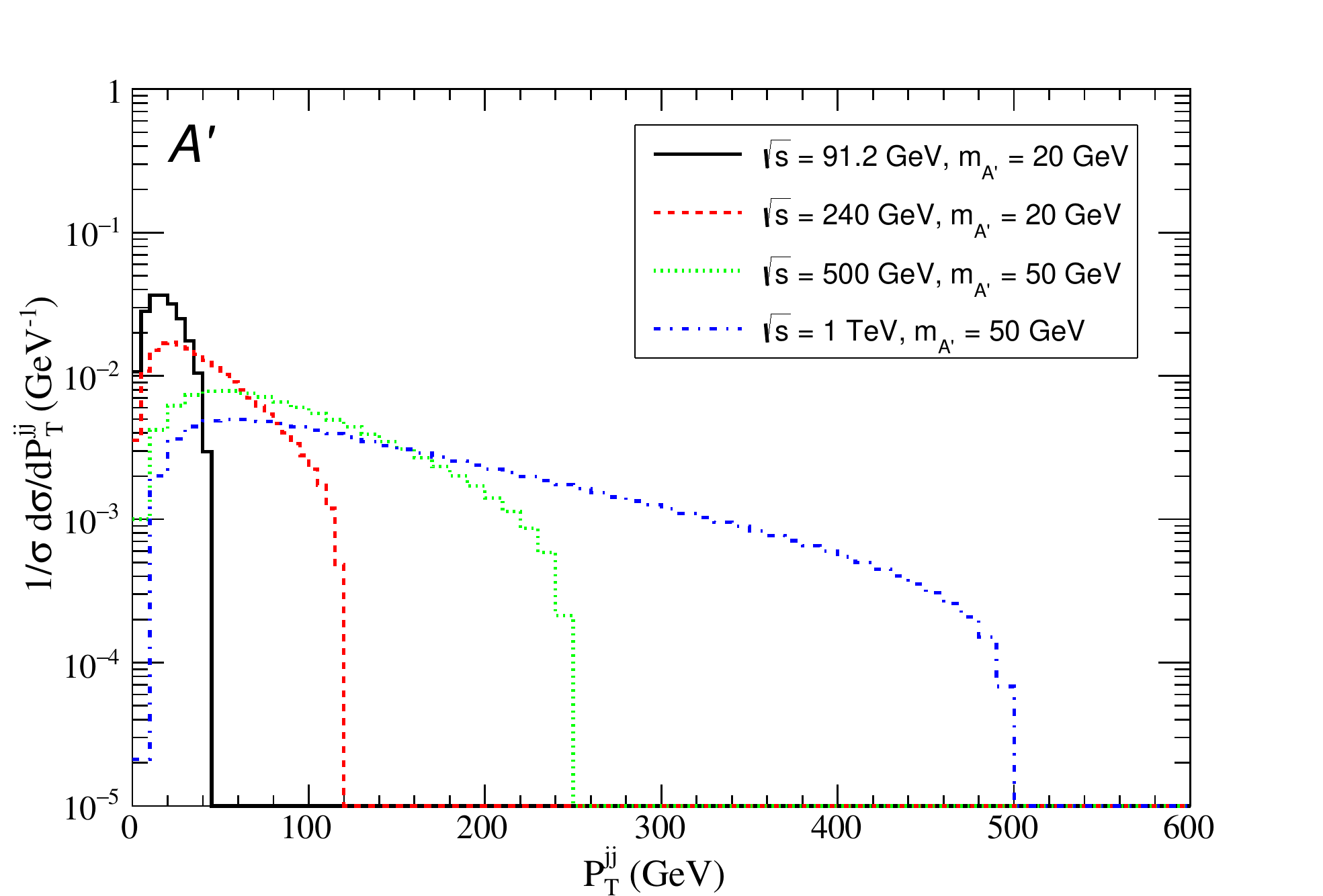}
\includegraphics[width=2.2in]{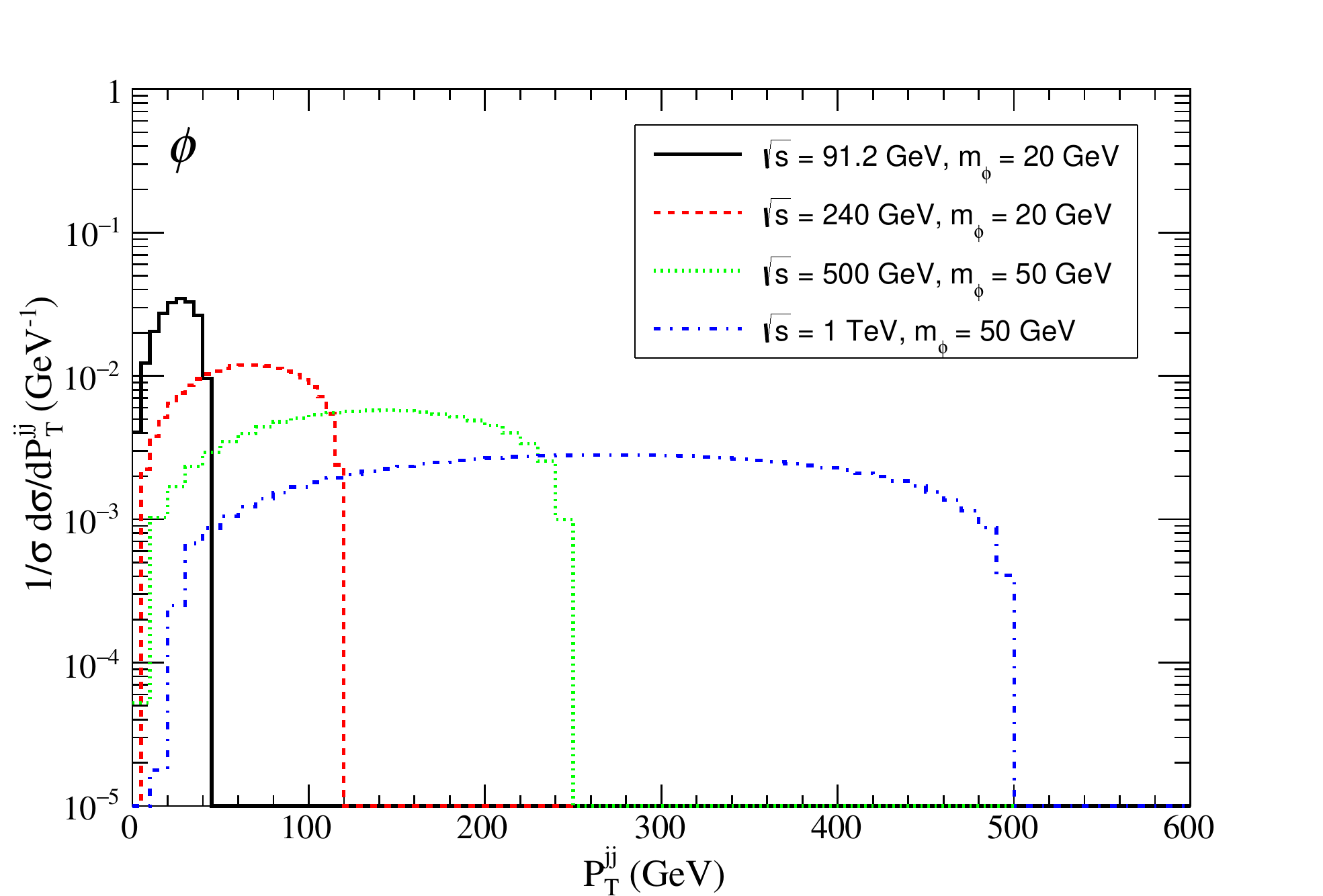}
\includegraphics[width=2.2in]{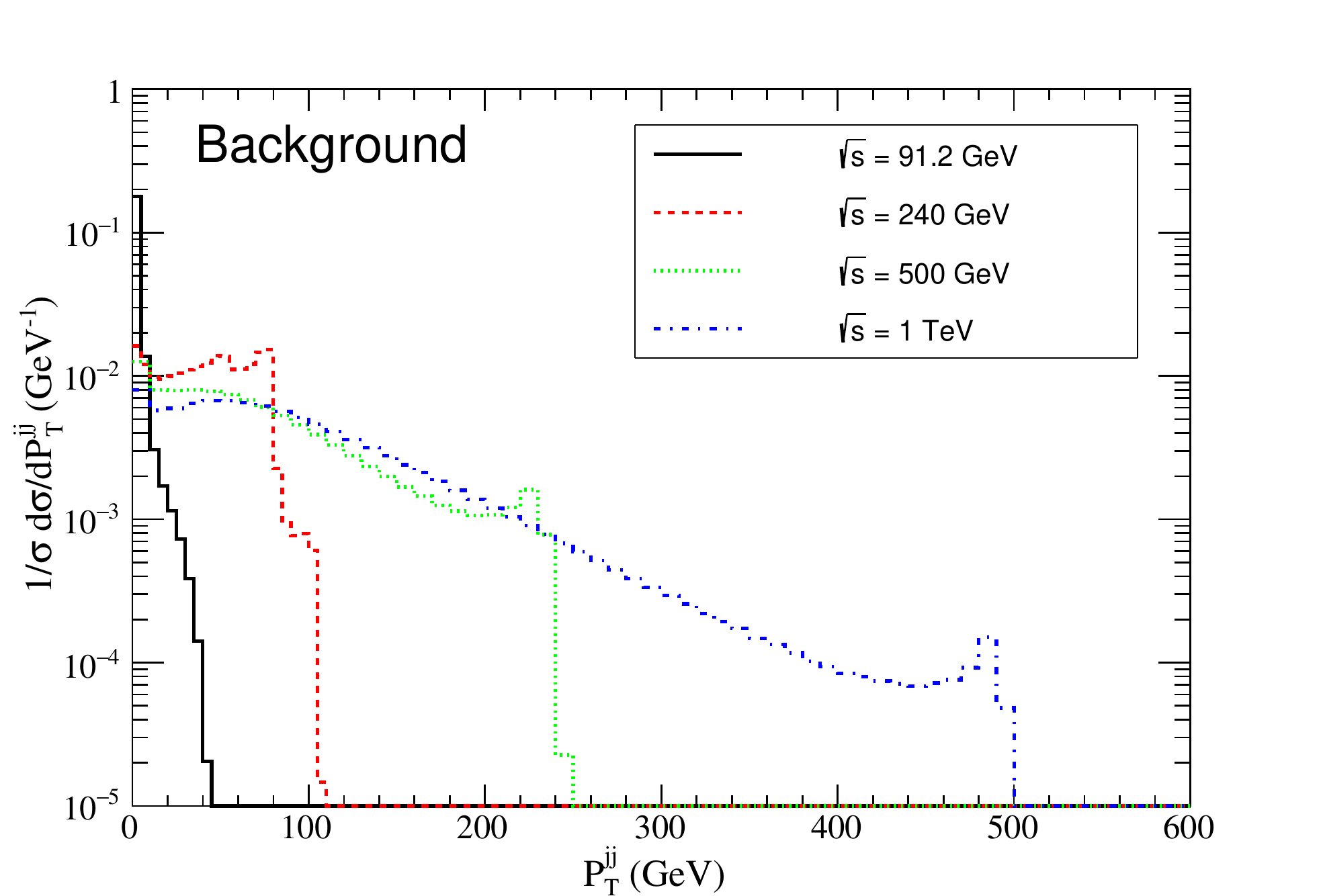}
\includegraphics[width=2.2in]{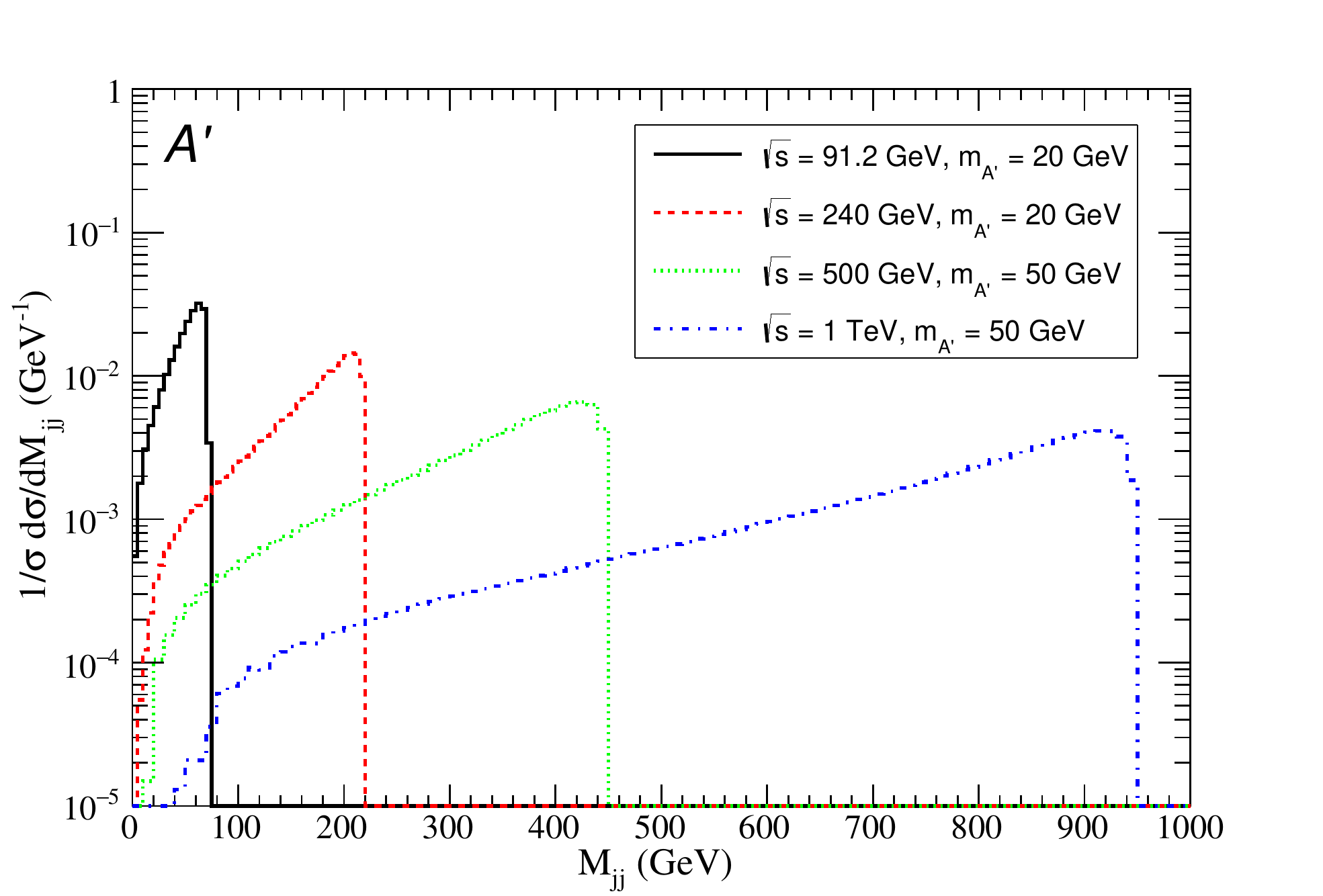}
\includegraphics[width=2.2in]{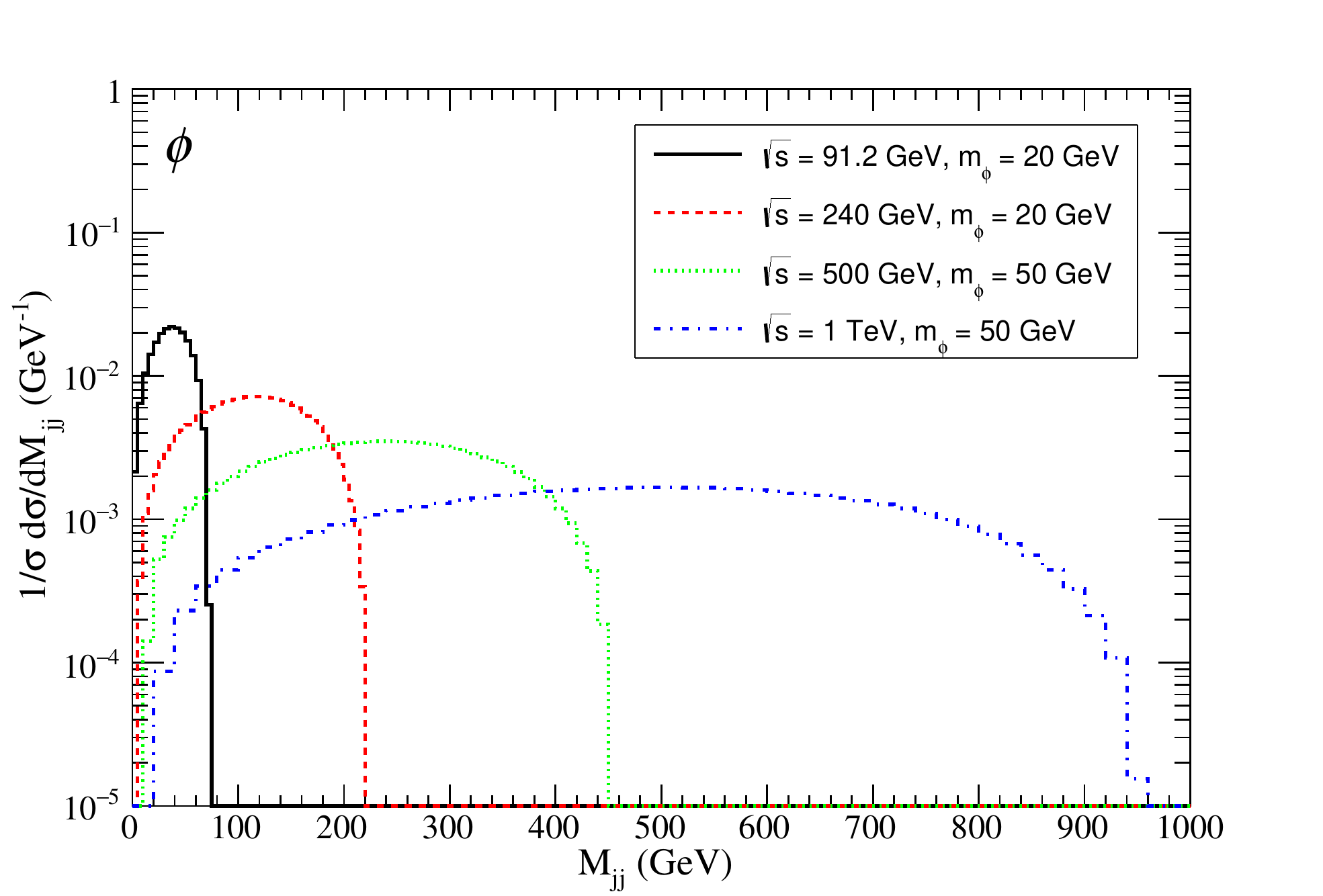}
\includegraphics[width=2.2in]{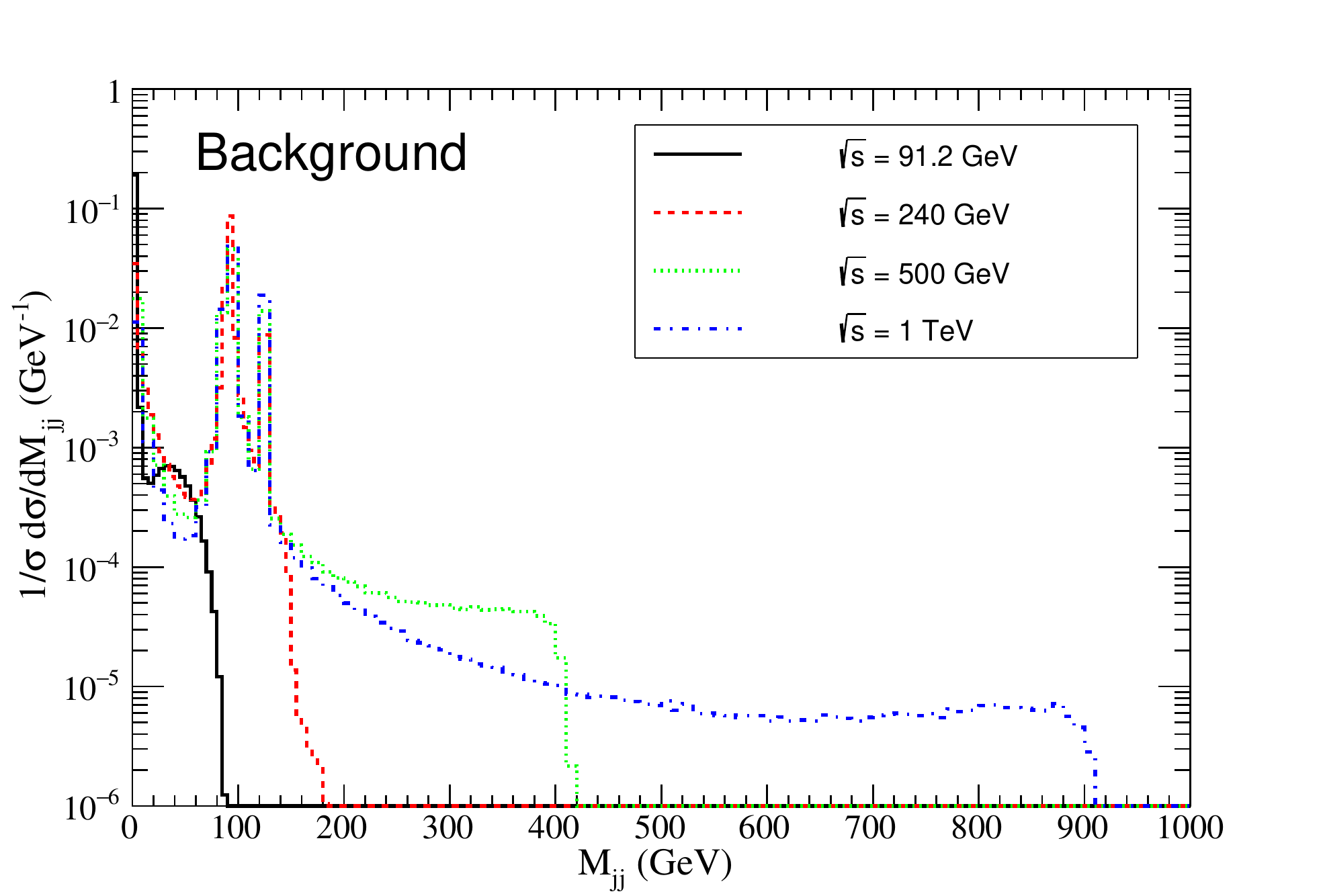}
\includegraphics[width=2.2in]{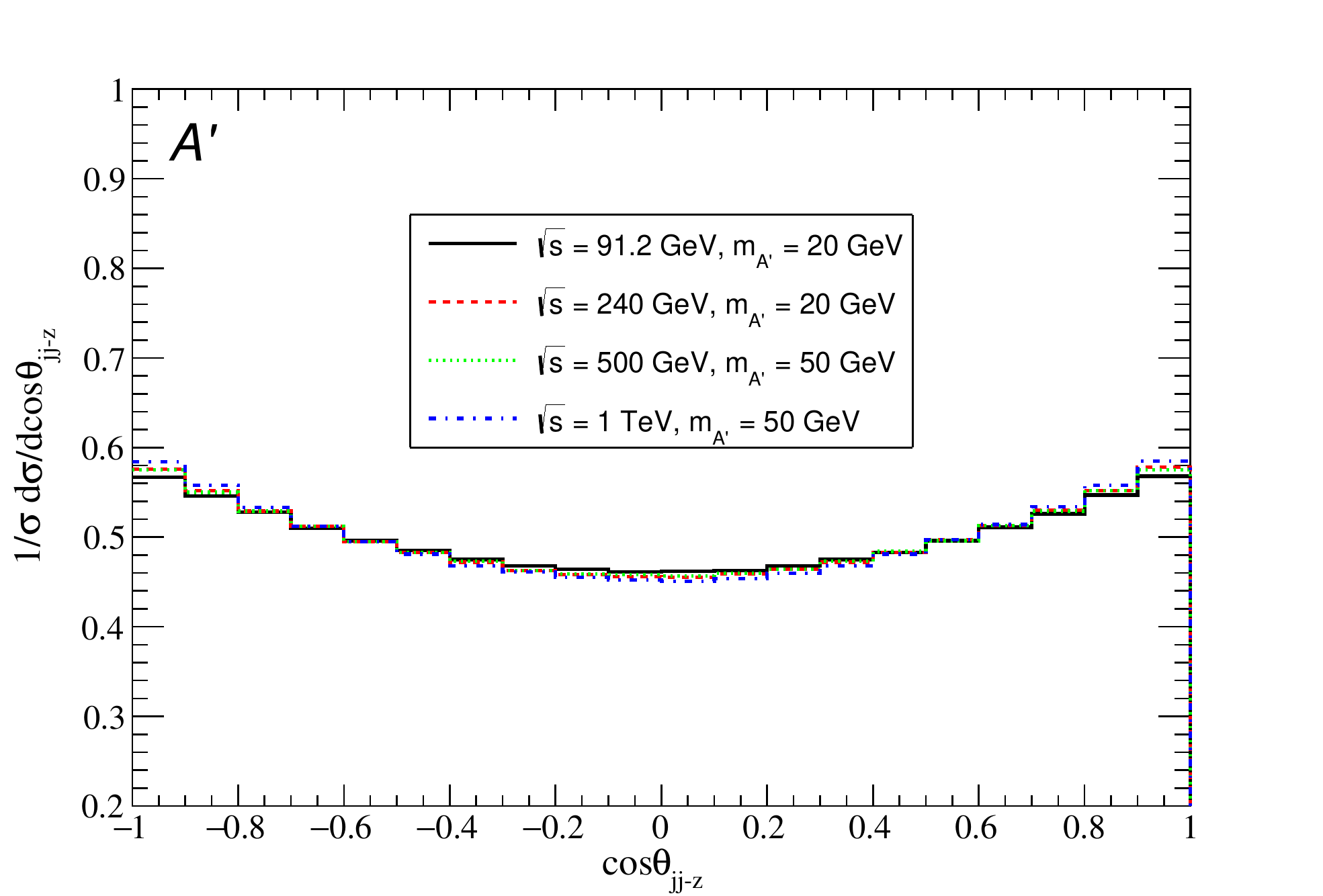}
\includegraphics[width=2.2in]{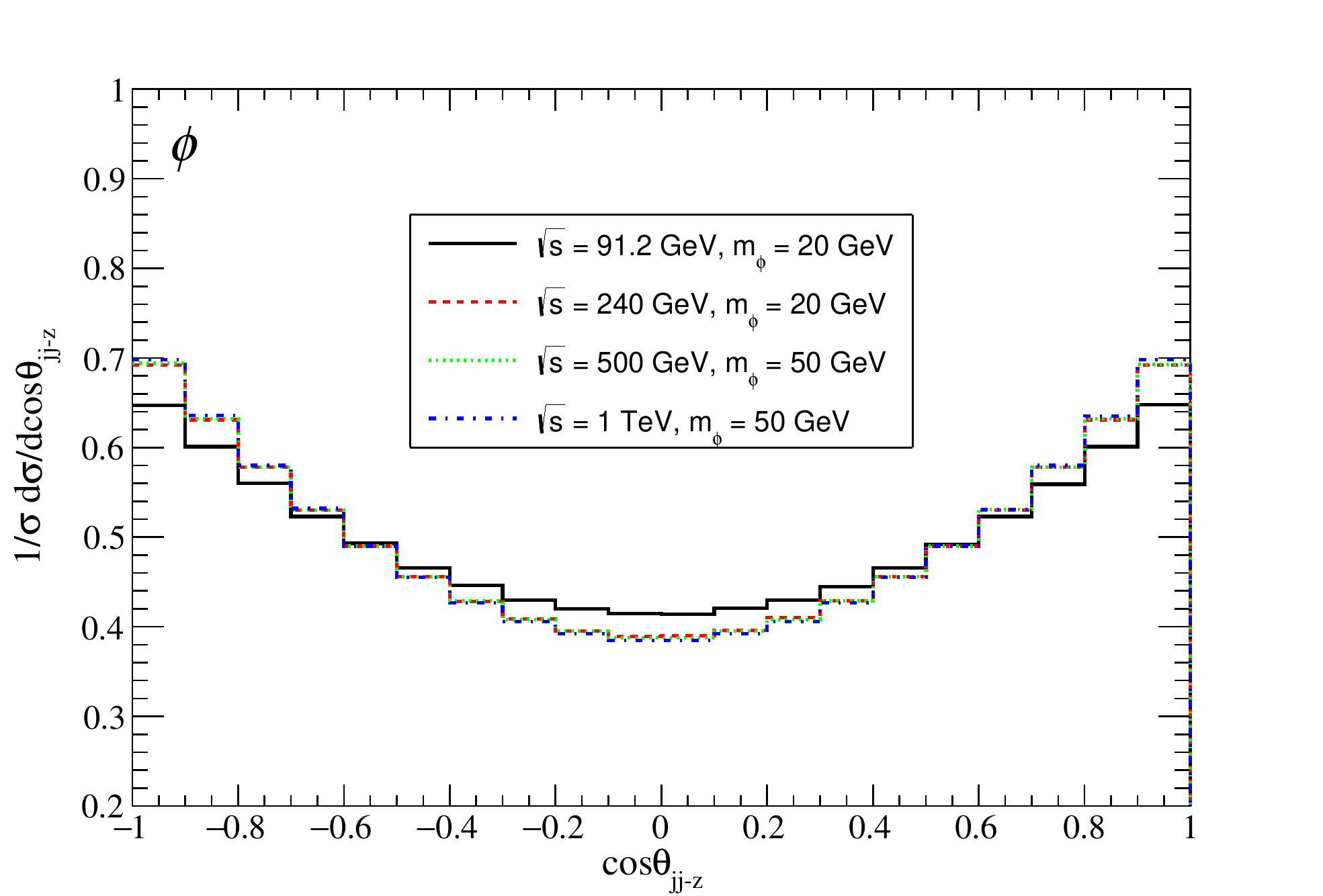}
\includegraphics[width=2.2in]{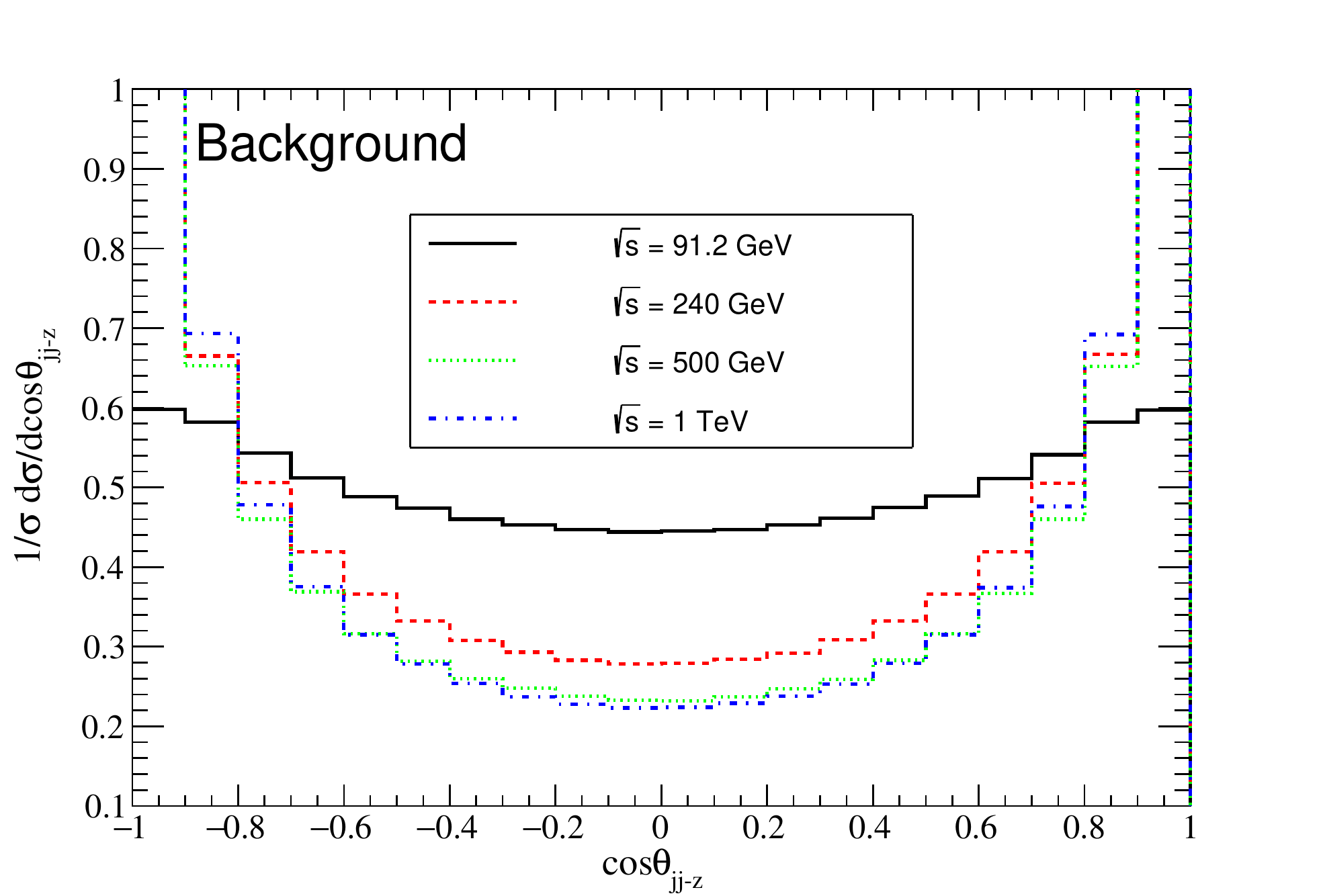}
\includegraphics[width=2.2in]{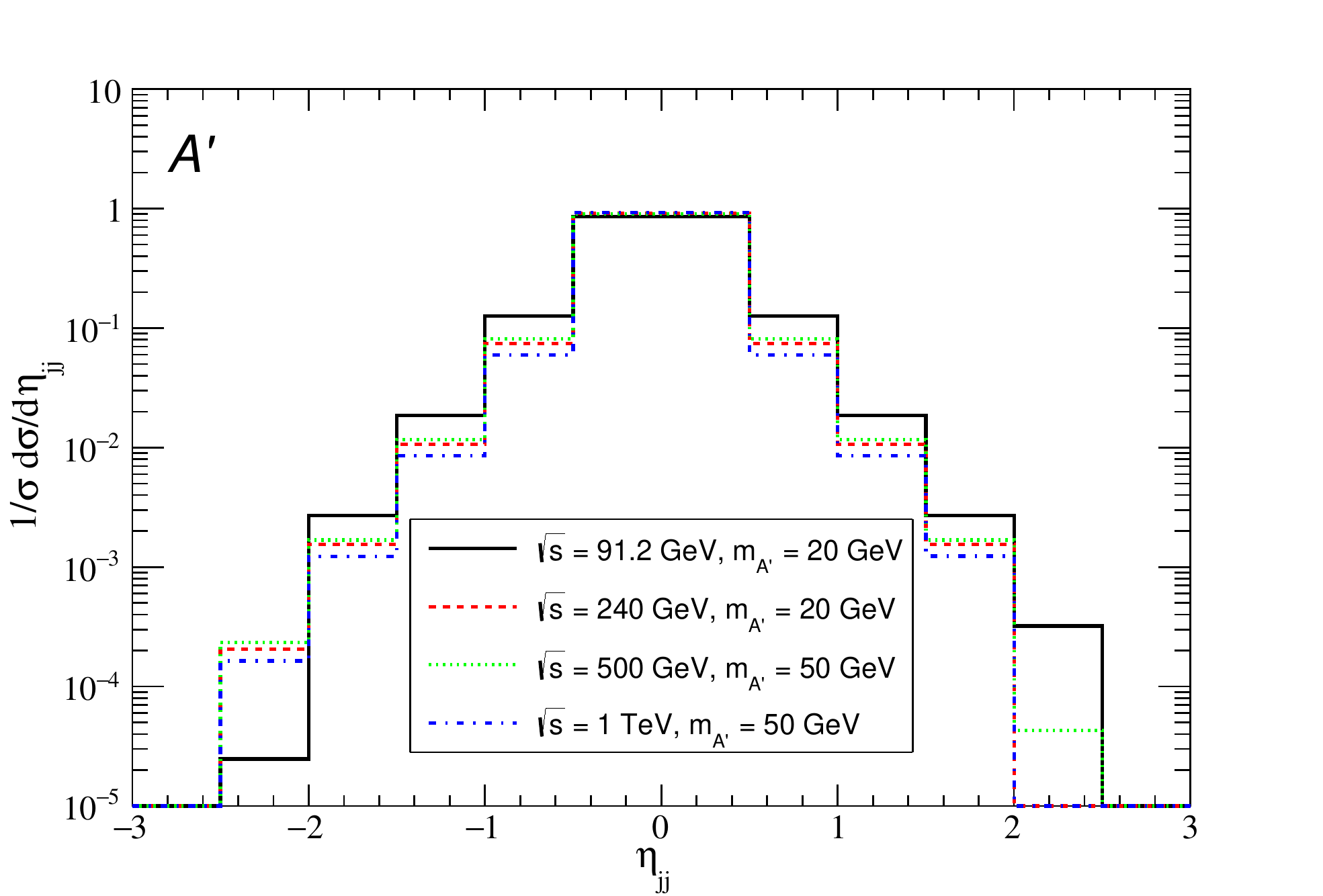}
\includegraphics[width=2.2in]{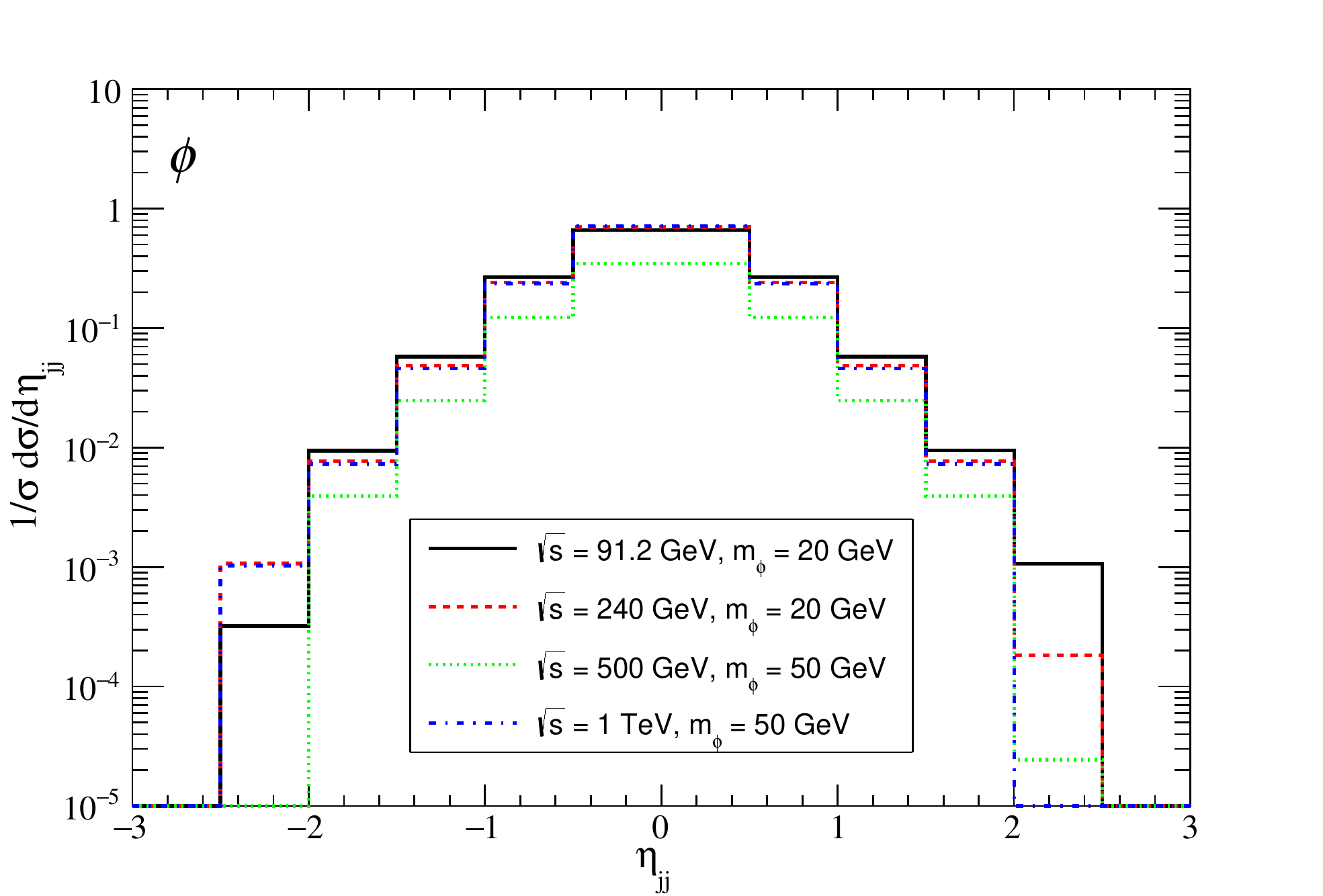}
\includegraphics[width=2.2in]{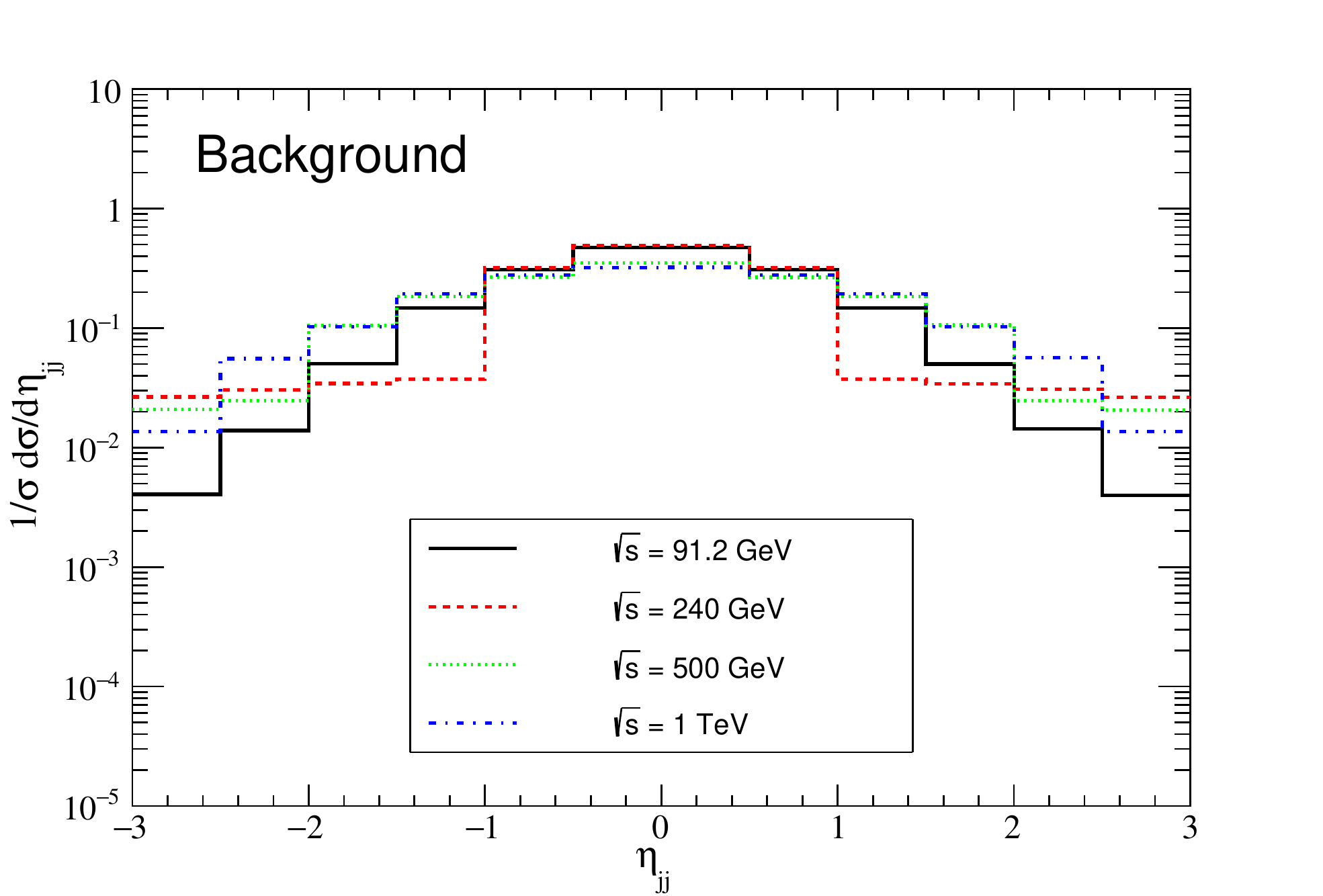}
\caption{Normalized $P_{T}^{jj}$, $M_{jj}$, $cos\theta_{jj-z}$ and $\eta_{jj}$ distributions of the two-jet system for the dark photon production process $e^{+} e^{-} \rightarrow q \bar{q} A^{\prime}$ (left panels), the dark scalar mediator production process $e^{+} e^{-} \rightarrow q \bar{q} \phi$ (middle panels), and the dominant background process $e^{+} e^{-} \rightarrow q\bar{q}\nu \bar \nu$ (right panels), for $\sqrt{s} = $ 91.2~GeV, 240~GeV, 500~GeV, and 1~TeV.}
\label{fig:kinematic_distributions_v_s_1} 
\end{figure}

\begin{figure}[htbp]
\centering
\includegraphics[width=3.2in]{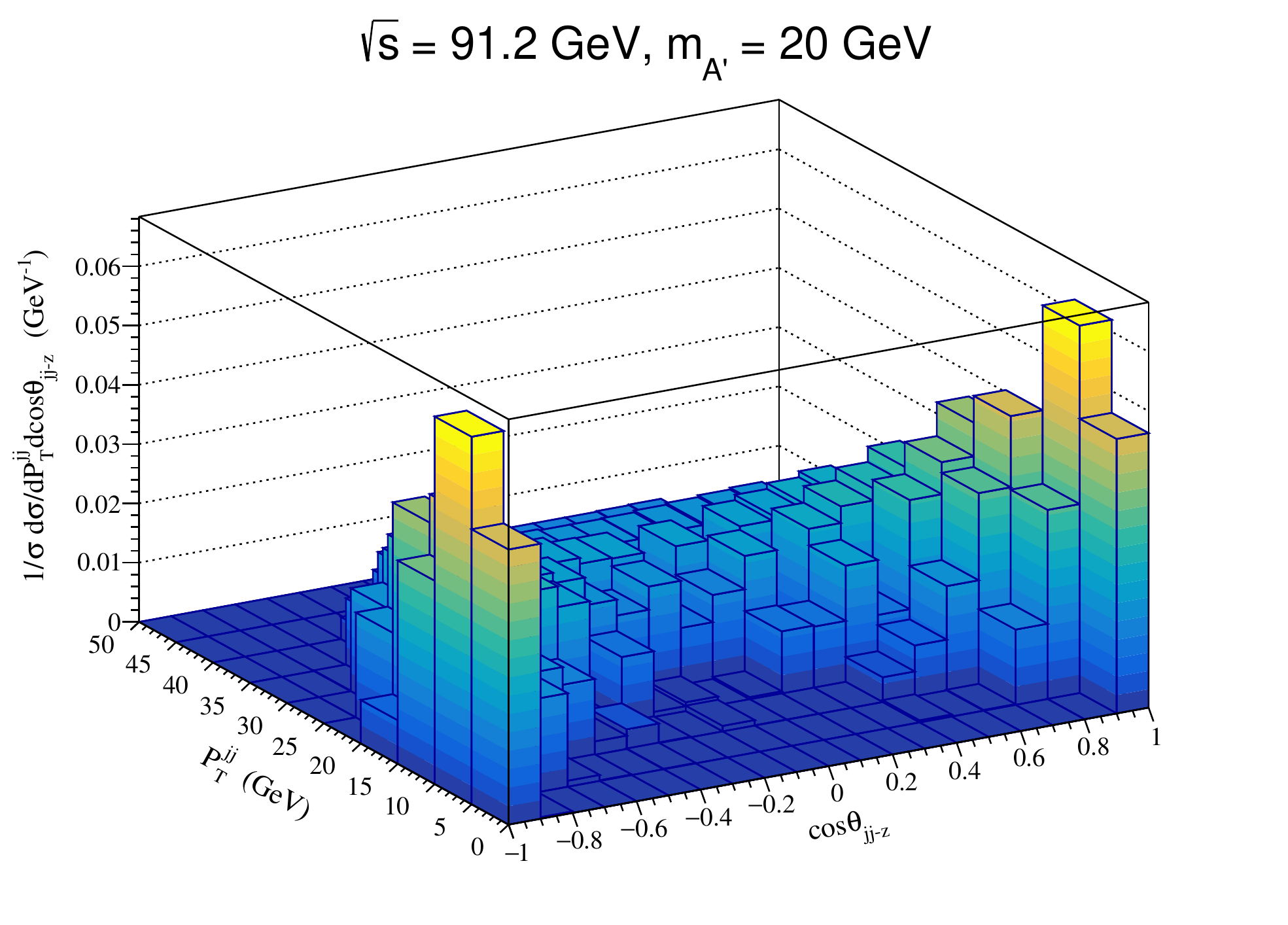}
\includegraphics[width=3.2in]{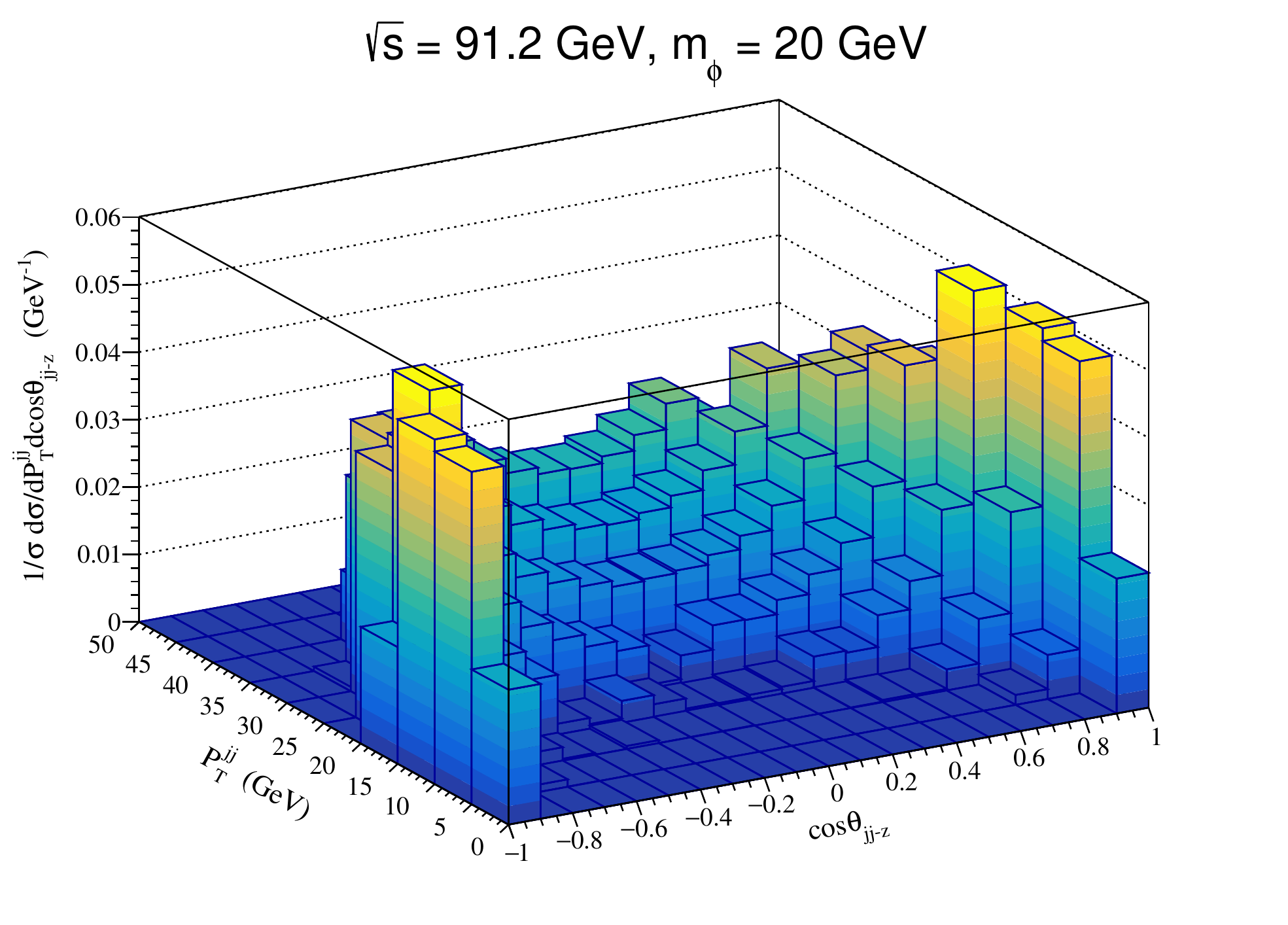}
\includegraphics[width=3.2in]{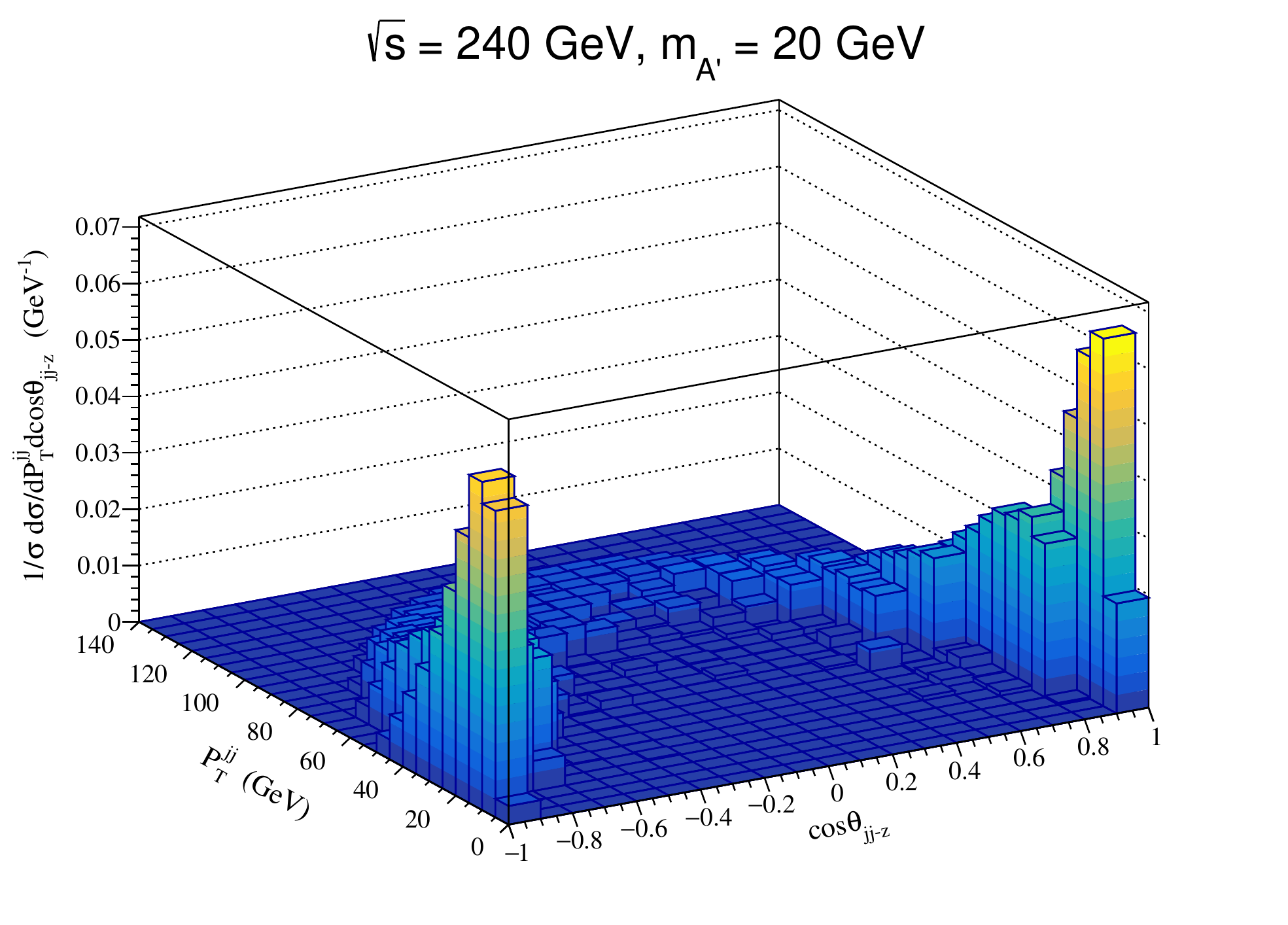}
\includegraphics[width=3.2in]{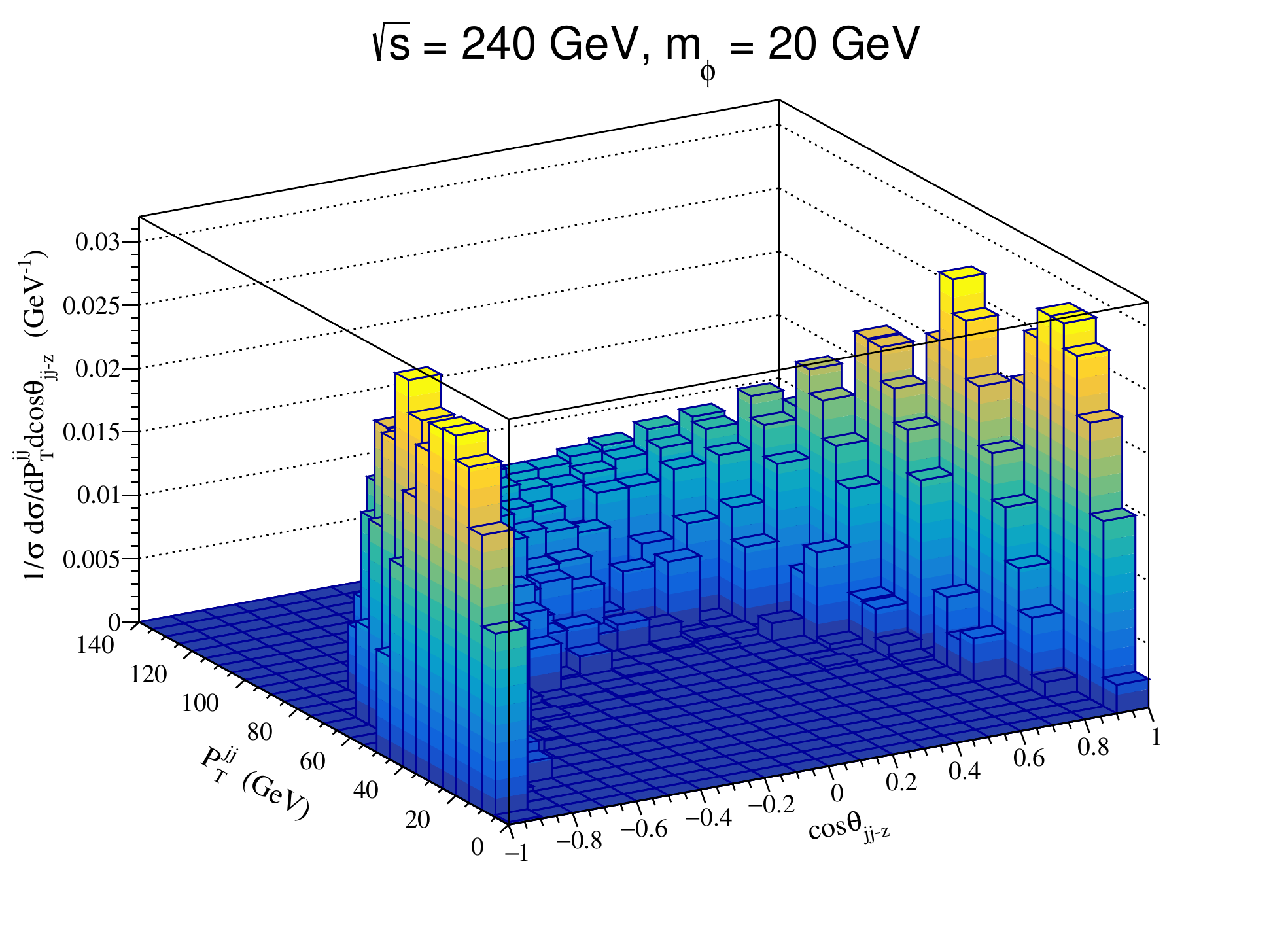}
\includegraphics[width=3.2in]{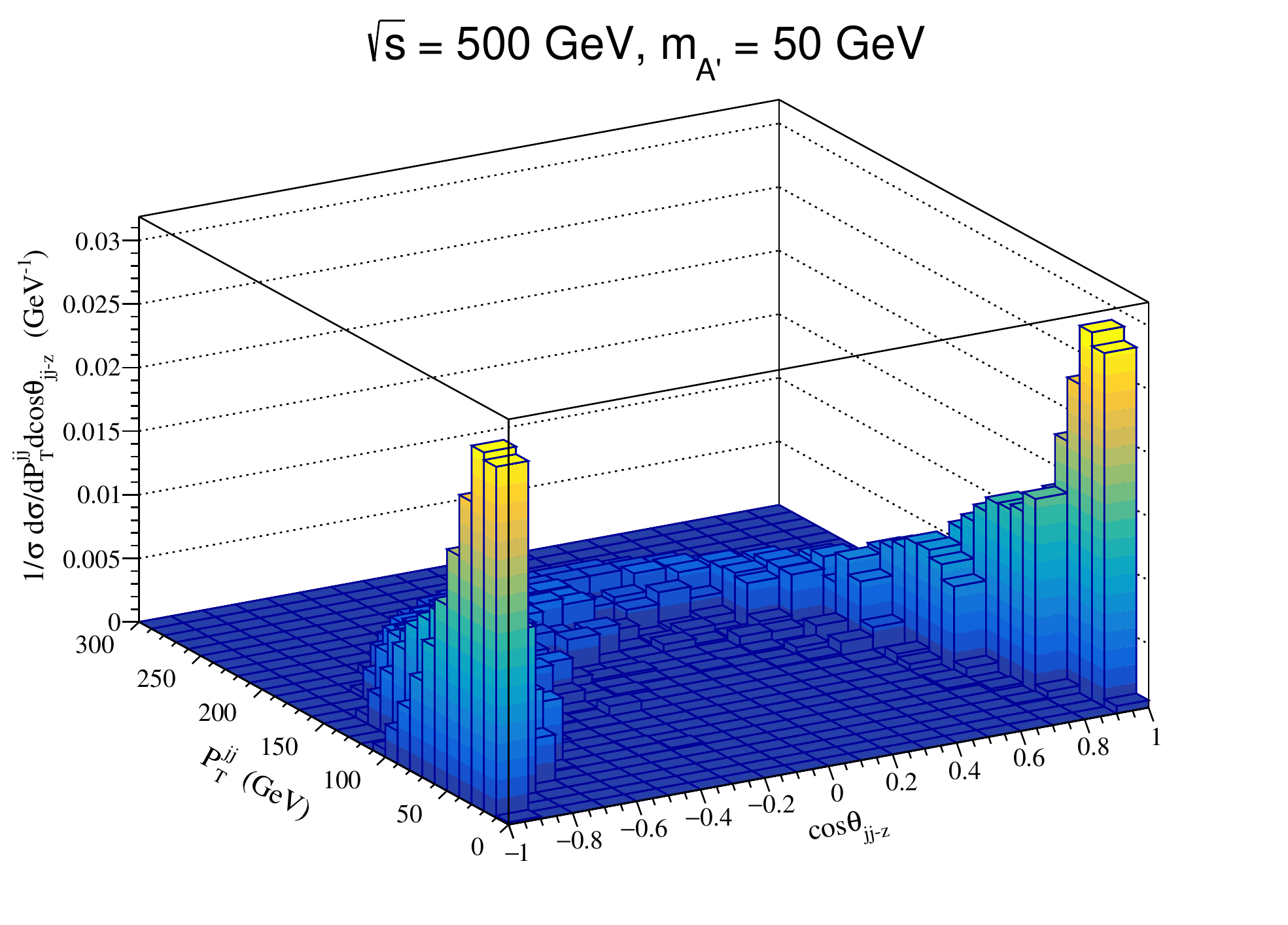}
\includegraphics[width=3.2in]{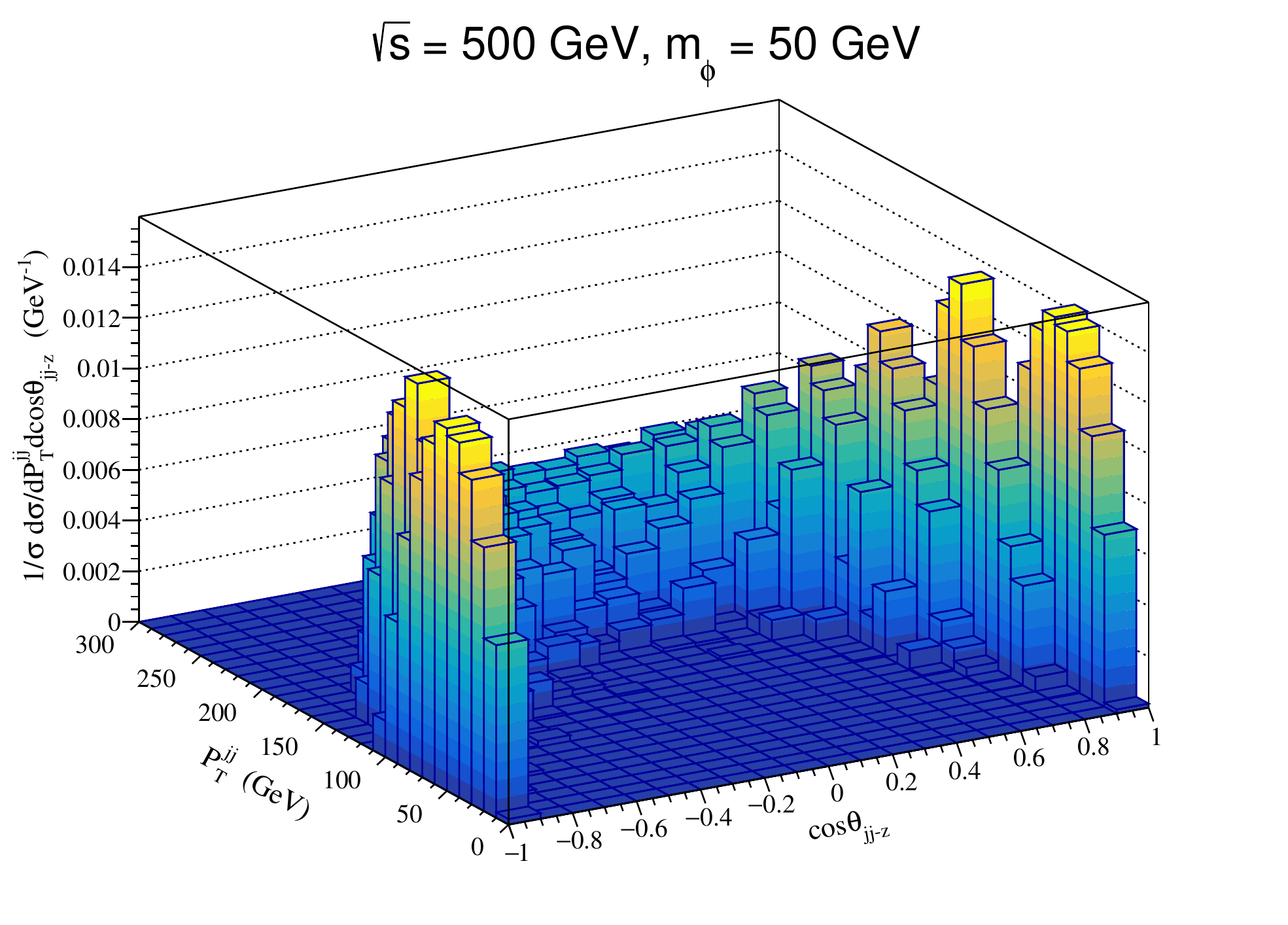}
\includegraphics[width=3.2in]{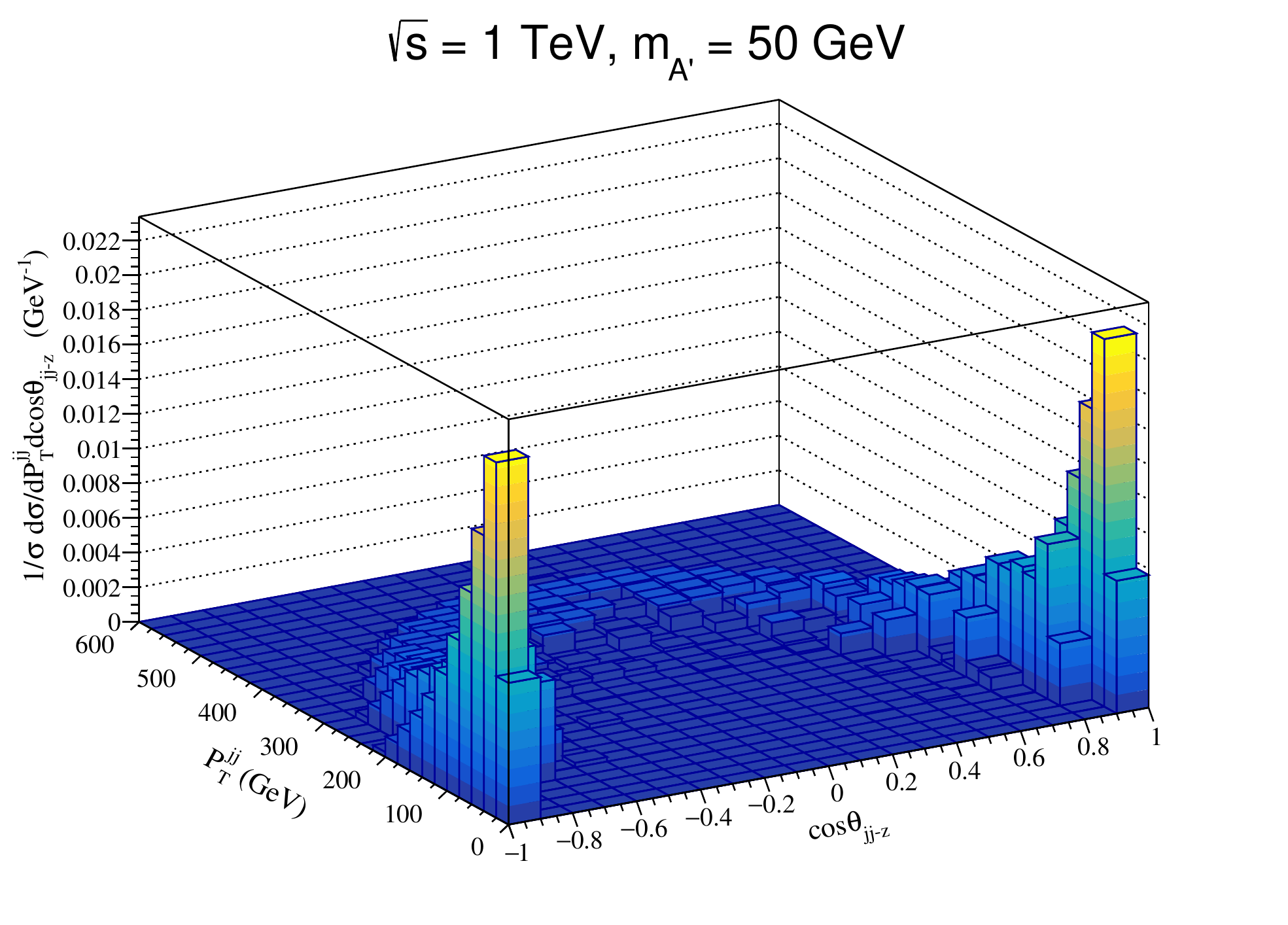}
\includegraphics[width=3.2in]{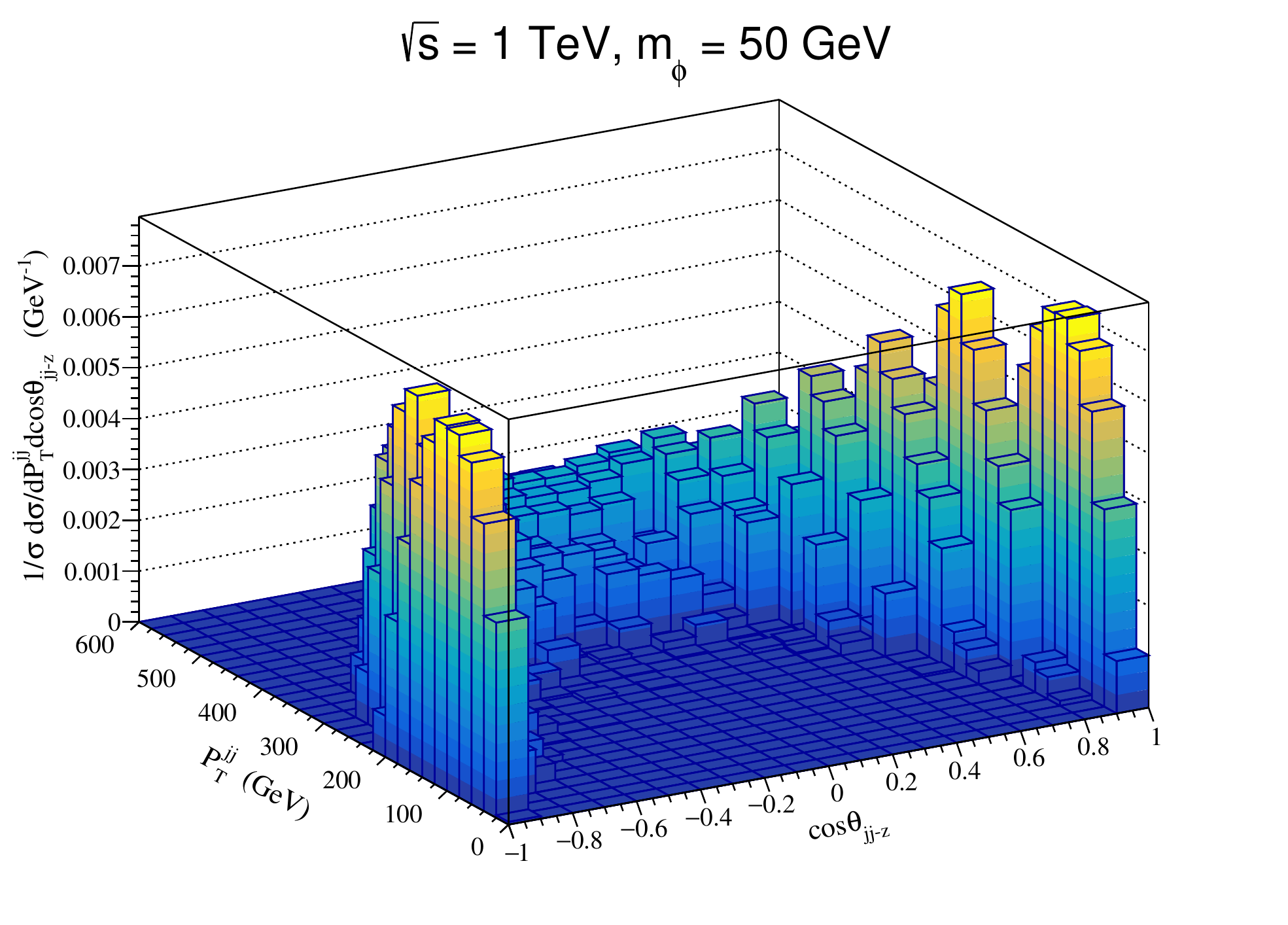}
\caption{Normalized distributions as function of $cos\theta_{jj-z}$ and $P_{T}^{jj}$ for the processes $e^{+} e^{-} \rightarrow q \bar{q} A^{\prime}$ (left panels) and $e^{+} e^{-} \rightarrow q \bar{q}\phi$ (right panels), for several $\sqrt{s}$ and $m_{A^{\prime}}$ ($m_{\phi}$) values without any kinematic cuts. }
\label{fig:kinematic_distributions_v_s_2} 
\end{figure}

 In order to obtain the analytical amplitudes, we use FeynArts~\cite{Hahn:2000kx} and FeynCalc~\cite{Shtabovenko:2016sxi} to generate the Feynman diagrams and perform the calculations. We use the multidimensional numerical integration package Cuba~\cite{Hahn:2016ktb} to analyze the kinematic distributions.
The cross-sections of the processes $e^{+} e^{-} \rightarrow q \bar{q} A^{\prime}$ ($e^{+} e^{-} \rightarrow q \bar{q} \phi$) are suppressed by factors of $\varepsilon^2$ ($\varepsilon_{s}^2$).
In order to see the general trend, we show the reduced cross-sections of the two processes as function of $\sqrt{s}$ and $m_{A^{\prime}}$ or $m_{\phi}$ in Fig.~\ref{fig:cross_section_v_s}. Fig.~\ref{fig:cross_section_v_s} (a) and (b) exhibit peaks due to the contribution from the resonant $Z^0$ boson production. Taking $m_{A^{\prime}} = 20$ GeV as an example, the cross-section decreases by about three orders of magnitude when $\sqrt{s}$ increases from 91.2 GeV to 1 TeV.
Fig.~\ref{fig:cross_section_v_s} (c) and (d) show that the reduced cross-sections become smaller as the mass becomes bigger. It is worth noting that since the value of the coupling parameter $\varepsilon~(\varepsilon_s$) varies with the mass $m_{A^{\prime}}~(m_{\phi})$, the shape of the cross-section changes with $m_{A^{\prime}}$ ($m_{\phi}$) when the mass dependent $\varepsilon$ ($\varepsilon_s$)is used. As we focus on the production of invisible dark photons $A^{\prime}$ and dark scalar mediators $\phi$ at $e^{+} e^{-}$ colliders, one can identify them by reconstructing the missing momentum, i.e. the recoil of two final jets.
  The four-momentum of the two-jet system is used to infer the characteristics of the two processes. Fig.~\ref{fig:kinematic_distributions_v_s_1} shows the normalized $P_{T}^{jj}$, $M_{jj}$, $cos\theta_{jj-z}$ and $\eta_{jj}$ distributions of the two-jet system for the processes $e^{+} e^{-} \rightarrow q\bar{q}A^{\prime}$ (left panels) and $e^{+} e^{-} \rightarrow q\bar{q} \phi$ (middle panels) for several $\sqrt{s}$ and $m_{A^{\prime}}$ ($m_{\phi}$) without any kinematic cuts. Here, $P_{T}^{jj}$ is the transverse momentum of the two-jet system and $M_{jj}$ is the invariant mass, $\theta_{jj-z}$ is the angle between the momentum of the two-jet system and the particle beam axis, and $\eta_{jj}$ is the rapidity of the two-jet system. For comparison, we use MadGraph~\cite{Alwall:2014hca} to analyze the kinematic distributions of the dominant background processes $e^{+} e^{-} \rightarrow q\bar{q} \nu \bar \nu$ ($\nu = \nu_{e},~\nu_{\mu}~\rm{and}~\nu_{\tau}$), which is shown in Fig.~\ref{fig:kinematic_distributions_v_s_1} (right panels). For $\sqrt{s} \geq 240~\rm{GeV}$, the $M_{jj}$ distributions of the background exhibit two peaks around $M_{jj}\approx$ 91~GeV and 125~GeV due to the contributions of the resonant $Z^0$ and the Higgs boson. However, for $\sqrt{s} = 91.2~\rm{GeV}$, the $Z^{0}$ peak is not obvious, because we have set the minimum transverse momentum of the jets to 0.5~GeV. From Fig.~\ref{fig:kinematic_distributions_v_s_1}, one can see that the kinematic distributions of the two processes are somewhat different. We further investigate these distributions as function of $cos\theta_{jj-z}$ and $P_{T}^{jj}$  in Fig.~\ref{fig:kinematic_distributions_v_s_2} for several $\sqrt{s}$ and $m_{A^{\prime}}$ ($m_{\phi}$) values.
Compared with the scalar mediator, the distributions for the dark photon $A^{\prime}$ are restricted to a smaller area. For example, for $\sqrt{s} = $91.2 GeV, the dominant area for $A^{\prime}$ is $cos\theta_{jj-z}\in (-1, -0.9) ~\rm{and} ~(0.9, 1)$ with $P_{T}^{jj}\in(0, 10)$, while the area for $\phi$ is comparatively broader. For higher center-of-mass energies $\sqrt{s}$, this trend is even more obvious.

As mentioned above, the kinematic distributions of the dark photon $A^{\prime}$ and dark scalar mediator $\phi$ are different. The difference can be enhanced by imposing appropriate kinematic cuts on the $P_{T}^{jj}$ and $cos\theta_{jj-z}$ distributions. As we show below, there are significant differences between the production of the dark photon and dark scalar mediator at $e^{+} e^{-}$ colliders.

\begin{figure}[htb]
\centering
\includegraphics[width=3.3in]{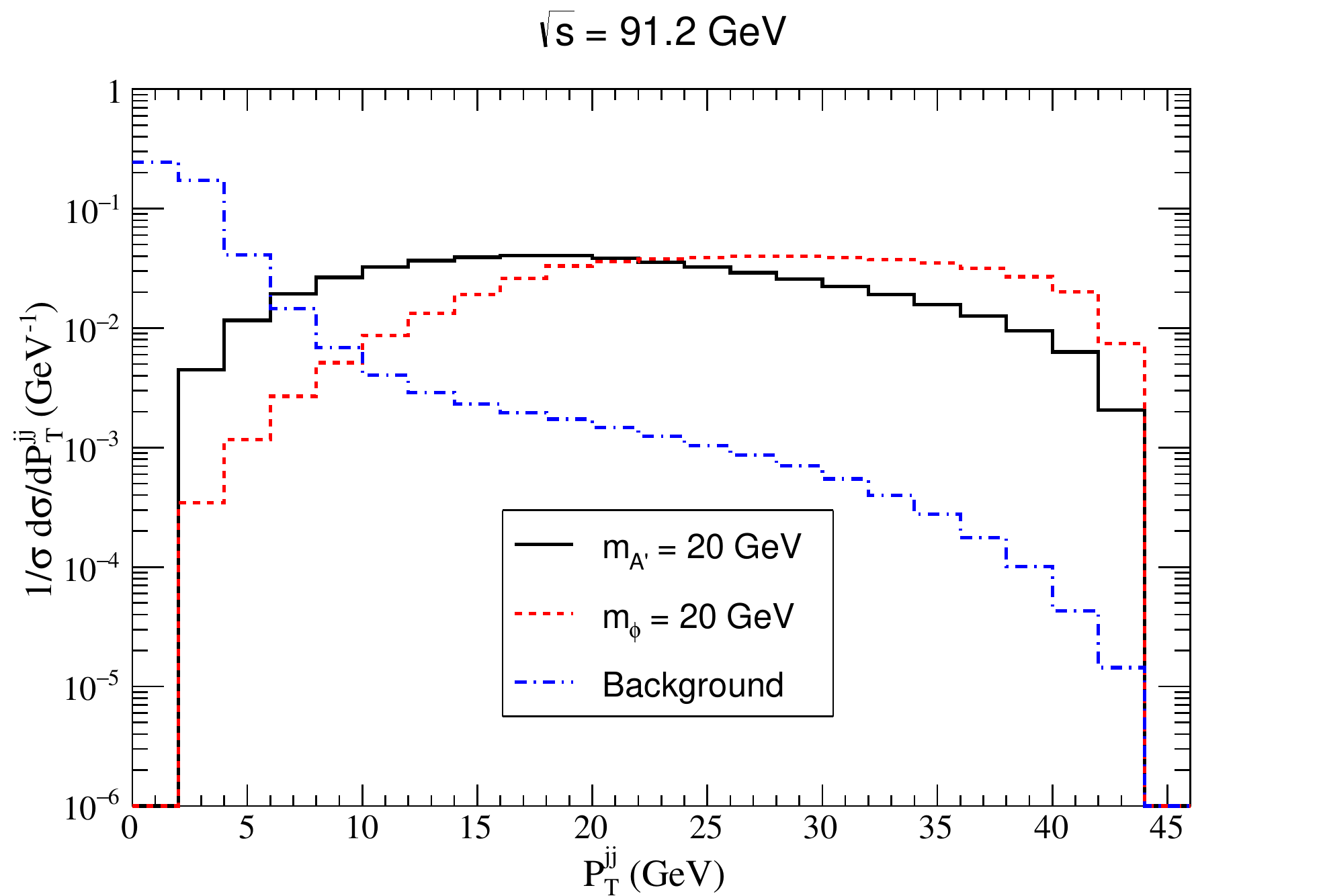}
\includegraphics[width=3.3in]{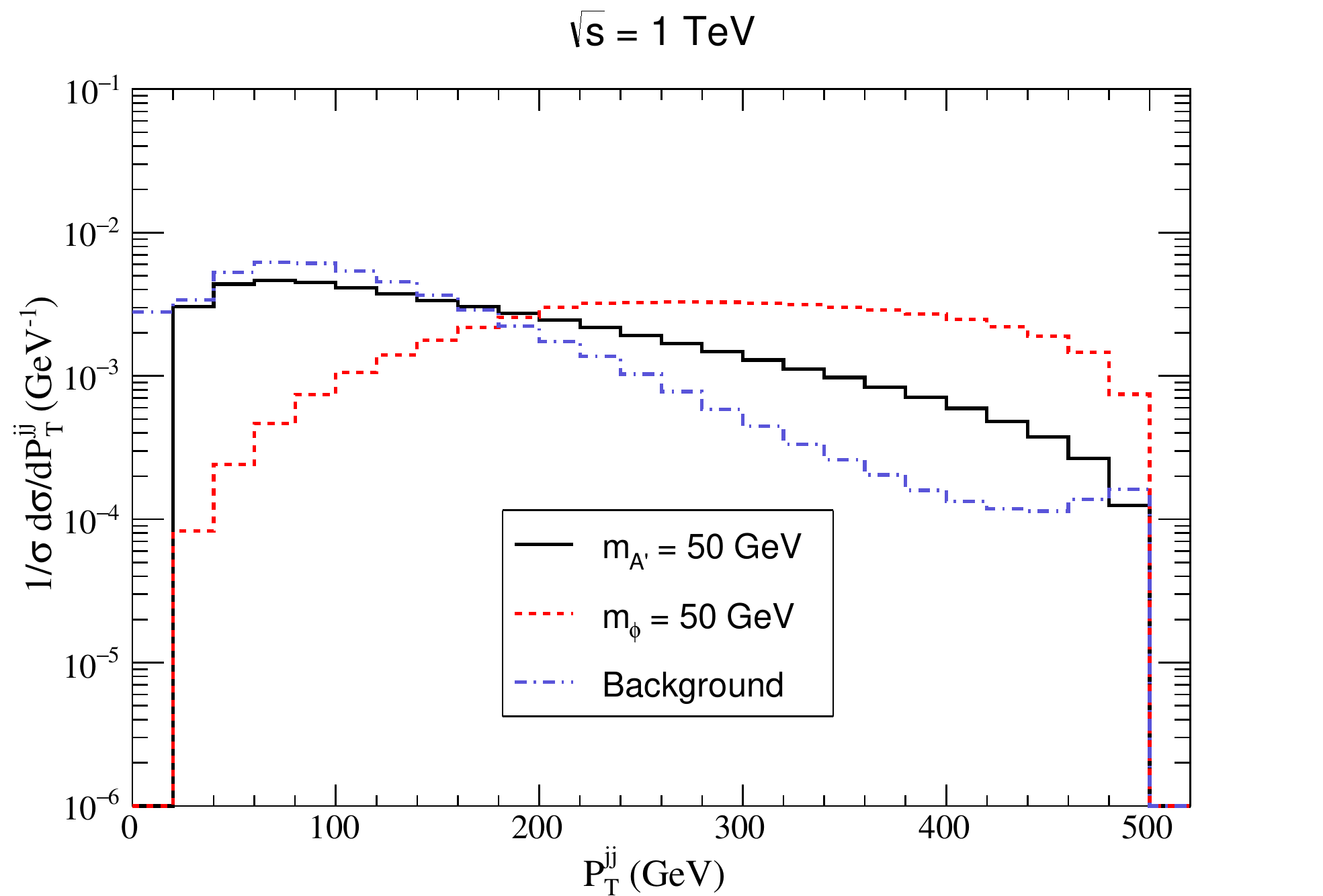}
\caption{Normalized $P_{T}^{jj}$ distributions of the two-jet system for the two processes and the background, for $\sqrt{s}$ = 91.2~GeV (left panel) and 1~TeV (right panel), with the cut $-0.9 < cos\theta_{jj_z} < 0.9$.}
\label{fig:vector_scalar_1} 
\end{figure}

In order to show the difference in $P_{T}^{jj}$, we impose a cut on $cos\theta_{jj-z}$ such that $-0.9 < cos\theta_{jj-z} < 0.9$, presented in Fig.~\ref{fig:vector_scalar_1}.
In the case of $\sqrt{s} = $ 91.2 GeV and $m_{A^{\prime}}(m_{\phi})$ = 20~GeV and for $18~\textrm{GeV} \lesssim P_{T}^{jj} \lesssim 40~\textrm{GeV}$, the $P_{T}^{jj}$ distributions of $e^{+} e^{-} \rightarrow q \bar{q} A^{\prime}$ are attenuated as $P_{T}^{jj}$ increases, while the $P_{T}^{jj}$ distributions of $e^{+} e^{-} \rightarrow q \bar{q} \phi$ are substantially flat in this region. In the case of $\sqrt{s} = $1 TeV and $m_{A^{\prime}}(m_{\phi})$ = 50 GeV, the differences of the $P_{T}^{jj}$ distributions of the two processes become much easier to identify. For $60~\textrm{GeV}\lesssim P_{T}^{jj} \lesssim 460~\textrm{GeV}$, the $P_{T}^{jj}$ distributions of $e^{+} e^{-} \rightarrow q \bar{q} A^{\prime}$ are monotonically attenuated as $P_{T}^{jj}$ increases. However, the $P_{T}^{jj}$ distributions of $e^{+} e^{-} \rightarrow q \bar{q} \phi$ first increase quickly, and then decrease slowly in the same region. We also display the transverse momentum distributions of the background with the same cuts, which show quite a different shape for $\sqrt{s}= 91.2~\rm{GeV}$.

\begin{figure}[!htb]
\centering
\includegraphics[width=2.2in]{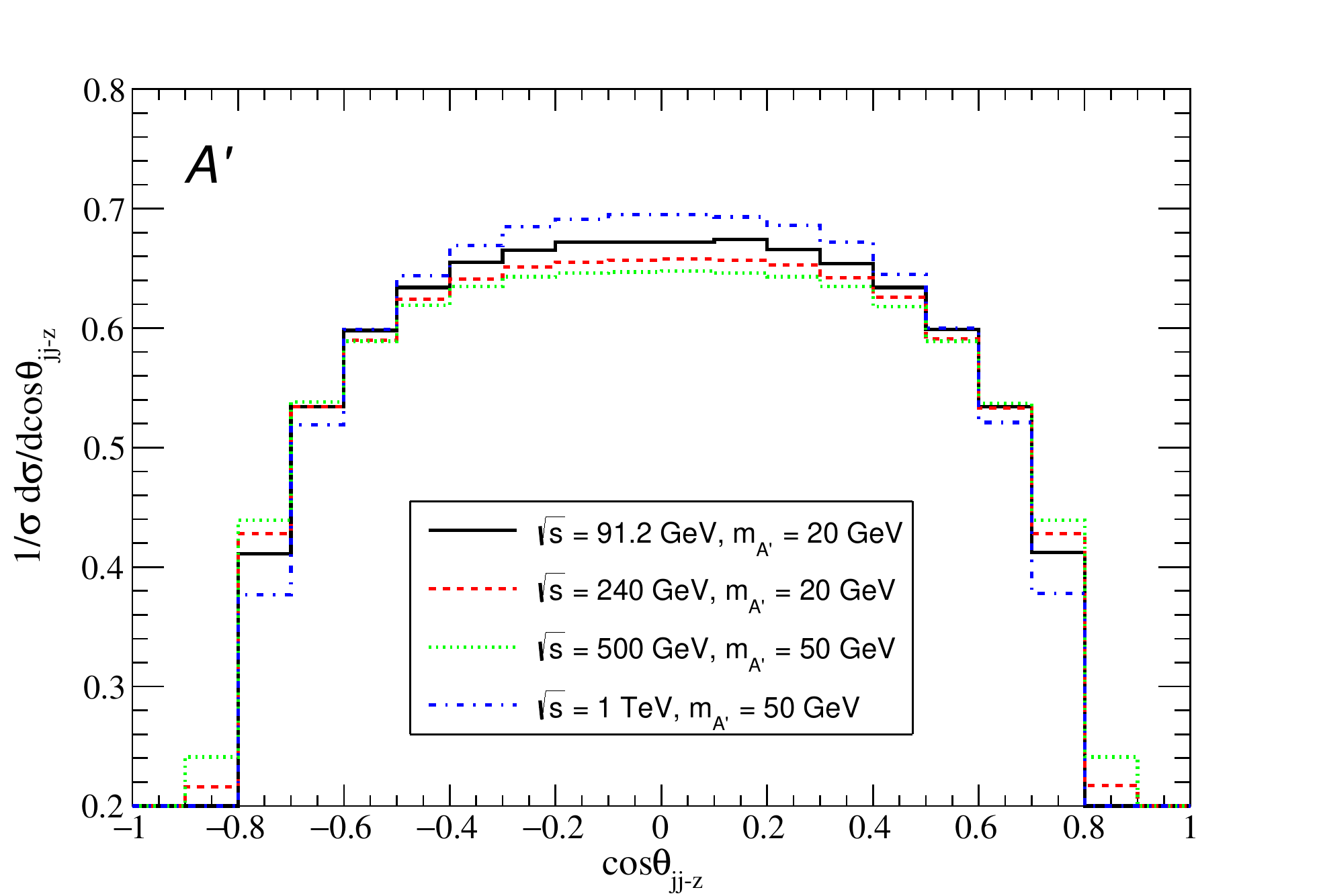}
\includegraphics[width=2.2in]{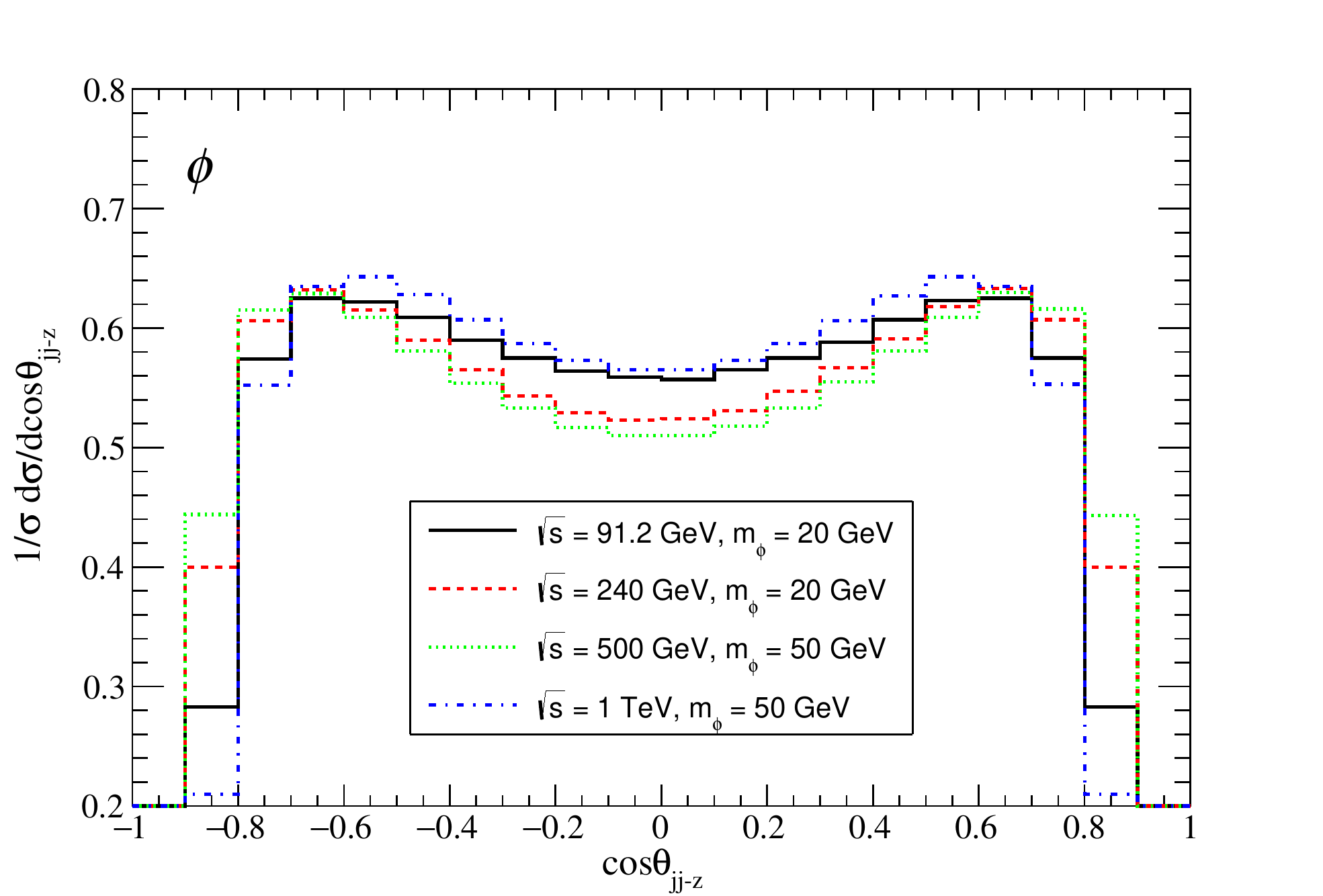}
\includegraphics[width=2.2in]{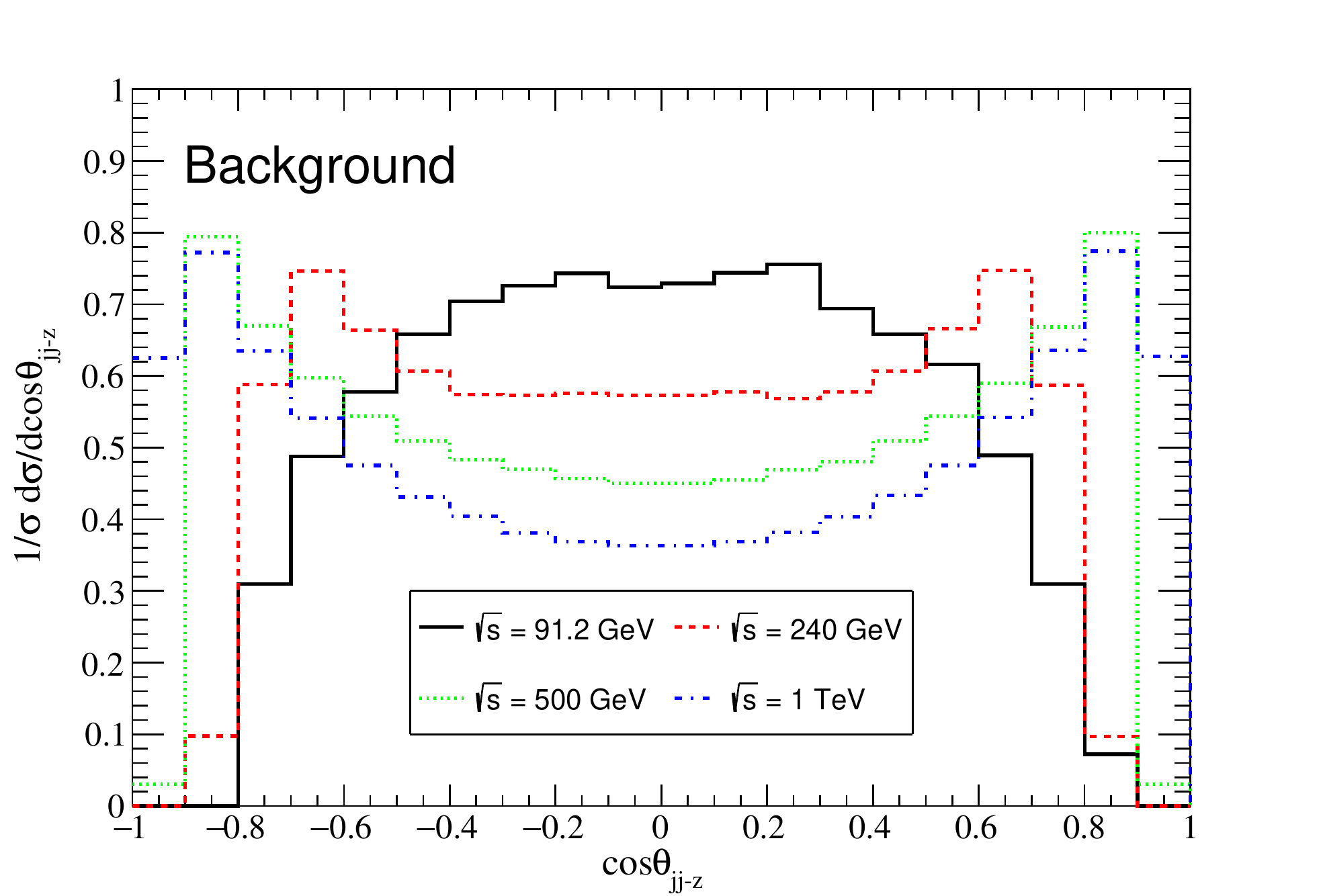}
\caption{Normalized $cos\theta_{jj}$ distributions of the two-jet system for the two processes and the background with cuts $P_{T}^{jj}>$ 20, 50, 100 and 240~GeV, for $\sqrt{s}$ = 91.2~GeV, 240~GeV, 500~GeV and 1~TeV.}
\label{fig:vector_scalar_2} 
\end{figure}

As examples of $cos\theta_{jj-z}$ distributions, we present in Fig. \ref{fig:vector_scalar_2} the differential cross-sections of the two processes for $\sqrt{s}$ = 91.2 GeV, 240 GeV, 500 GeV, 1 TeV and $m_{A^{\prime}}(m_{\phi})$ = 20 GeV or 50 GeV. It can be seen that the cuts of $P_{T}^{jj}$ can enhance the difference between the dark photon and dark scalar mediator.
Imposing the cuts $P_{T}^{jj}>$ 20, 50, 100 and 240~GeV for the above center-of-mass energies, we find that the differential distributions of $e^{+} e^{-}\rightarrow q \bar{q} A^{\prime}$ reach a maximum around $cos \theta_{jj}= 0 $, with the inverted  ``U" shape. However, for the scalar mediator, the maximum of the peak lies around $cos \theta_{jj}= \pm 0.7$, and the shape looks like the letter ``M". The shape of the angular distribution of the background with the same cuts varies dramatically for typical $\sqrt{s}$ values.

\newpage
\section{Identifying the dark photon signal against the background}

Future $e^{+} e^{-}$ colliders are expected to play a crucial role in discovering the nature of DM (dark sector) particles since they have a cleaner background.
In this section, we focus on how to identify the heavy dark photon $A^{\prime}$ signal against the expected background at a future CEPC experiment. The analysis is similar for the dark scalar mediator $\phi$.
In the dark photon model of Eq.~(\ref{Lagrangian_v}), $A^{\prime}$ can decay into a pair of SM quarks and a DM pair. The related decay widths are defined as
\begin{eqnarray}
\Gamma(A^{\prime}\rightarrow\chi \bar{\chi})&=&\frac{g_{\chi}^{2}(m_{A^{\prime}}^{2}+2m_{\chi}^{2})\sqrt{m_{A^{\prime}}^{2}-4m_{\chi}^{2}}}{12\pi m_{A^{\prime}}^{2}},\nonumber\\
\sum\limits_{q}\Gamma(A^{\prime}\rightarrow q \bar{q})&=&\sum\limits_{q}\frac{\varepsilon^{2}e^{2}c_{q}^{2}(m_{A^{\prime}}^{2}+2m_{q}^{2})\sqrt{m_{A^{\prime}}^{2}-4m_{q}^{2}}}{4\pi m_{A^{\prime}}^{2}},
\end{eqnarray}
where $c_{q}$ is the charge of the quarks. The branching ratio of $A^{\prime}\rightarrow\chi \bar{\chi}$ can be written as
\begin{eqnarray}
\textrm{Br}(A^{\prime}\rightarrow\chi \bar{\chi})=\frac{\Gamma(A^{\prime}\rightarrow\chi \bar{\chi})}{\Gamma(A^{\prime}\rightarrow\chi \bar{\chi})+\sum\limits_{q}(A^{\prime}\rightarrow q \bar{q})},
\end{eqnarray}
which is related to $g_{\chi}$ and $\varepsilon$, while the combined parameter $\alpha_{\chi}\varepsilon^{2}$ can be obtained from  Fig.1. Here we choose $m_{\chi}$ = 8.6~GeV, $g_{\chi}=0.032$, we extract $\varepsilon$ from the XENON-1T curve in Fig.~\ref{fig:upper_limit_of_epsilon}, and obtain the branching ratios of $A^{\prime}\rightarrow \chi \bar{\chi}$ listed in Table.~\ref{epsilon_value}. In the following, we study the $e^{+} e^{-}\rightarrow q \bar{q} A^{\prime}$ process with $A^{\prime}\rightarrow \chi \bar{\chi} $ due to its cleaner background.
 \begin{table}[htbp]
 \caption{The mixing parameter $\varepsilon$ and the branching ratios of $A^{\prime}\rightarrow\chi \bar{\chi}$ as function of the dark photon mass $m_{A^{\prime}}$, for $m_{\chi}$ of 8.6~GeV and $g_{\chi}$ of 0.032.}
\centering
\newcolumntype{d}{D{.}{.}{2}}
\begin{tabular}{cccccc}
  \hline
  $m_{A^{\prime}}$ &20 GeV& 30 GeV & 40 GeV & 50 GeV & 60 GeV \\ \hline
  $\varepsilon$ &  0.0030 & 0.0067 & 0.012 & 0.019&0.027  \\
  Br$(A^{\prime}\rightarrow\chi \bar{\chi})$ &  0.996 & 0.985 & 0.955 & 0.898& 0.809  \\ \hline
\end{tabular}
\label{epsilon_value}
\end{table}
The dominant background process is $e^{+} e^{-}\rightarrow q \bar{q} \nu \bar \nu$ ($\nu = \nu_{e},\nu_{\mu}$, and~$\nu_{\tau}$). In the final states of both the signal and background processes, we observe only two jets. The background process is simulated by MadGraph~\cite{Alwall:2014hca}. The invariant mass $M_{RA^{\prime}}$ of the dark photon can be reconstructed from the recoil four-momentum of the two-jet system, where $M_{RA^{\prime}}$ is defined as,
\begin{eqnarray}
M_{RA^{\prime}} = \sqrt{(p_{e^{+}}+p_{e^{-}}-p_{j1}-p_{j2})^{2}},
 \end{eqnarray}
where $p_{e^{+}}$, $p_{e^{-}}$, $p_{j1}$ and $p_{j2}$ are the four-momenta of the incoming electron, positron and the two jets in the final states, respectively. We focus on the light quark jets ($q=$ $u$, $d$, $s$, $c$ and $b$) since the top quark decays quickly.

Theoretically, the on-shell dark photon events can be reconstructed precisely at $M_{RA^{\prime}} = m_{A^{\prime}}$ in the invariant mass spectrum. However, the detector has a finite energy resolution, which results in bump structures in the $M_{RA^{\prime}}$ spectrum. To make our estimate more realistic, we simulate this effect by smearing the jet energies assuming a Gaussian resolution,
\begin{eqnarray}
\frac{\delta (E)}{E} = \frac{A}{\sqrt{E}}\oplus B,
\end{eqnarray}\\
where $\delta (E)/E$ is the energy resolution, $A$ is the sampling term, $B$ a constant term, and $\oplus$ denotes the sum in quadrature. According to the CEPC CDR ~\cite{CEPCStudyGroup:2018ghi}, the energy resolution for light jets ranges from 6\% at $E=$20~GeV to 3.6\% at $E=$100~GeV. We adopt the parameters $A= 25.7\%$ and $ B= 2.4\%$. The smearing effect is introduced in the same way in the reconstruction of the background events.

In order to identify the dark photon signal against the background, we need to impose proper kinematic cuts.
The cuts are based on the kinematic distributions of the signal and background processes.
We set the basic transverse momentum cut at $P_{T} >$ 10 GeV and the rapidity cut at $|\eta_{j}| <$ 4. In order to identify an isolated jet, the angular distribution between jets $i$ and $j$ is defined by
\begin{eqnarray}
\triangle R_{ij}=\sqrt{\triangle\phi_{ij}^{2}+\triangle\eta_{ij}^{2}},
\end{eqnarray}
where $\triangle \phi_{ij}^{2}$ ($\triangle \eta_{ij}^{2}$) denotes the azimuthal angle (rapidity) difference between the two jets. In the two-jet system, we set the basic cut at $\triangle R > 0.4$ for both the signal and background processes.

\begin{figure}[htb]
\centering
\includegraphics[width=3.3in]{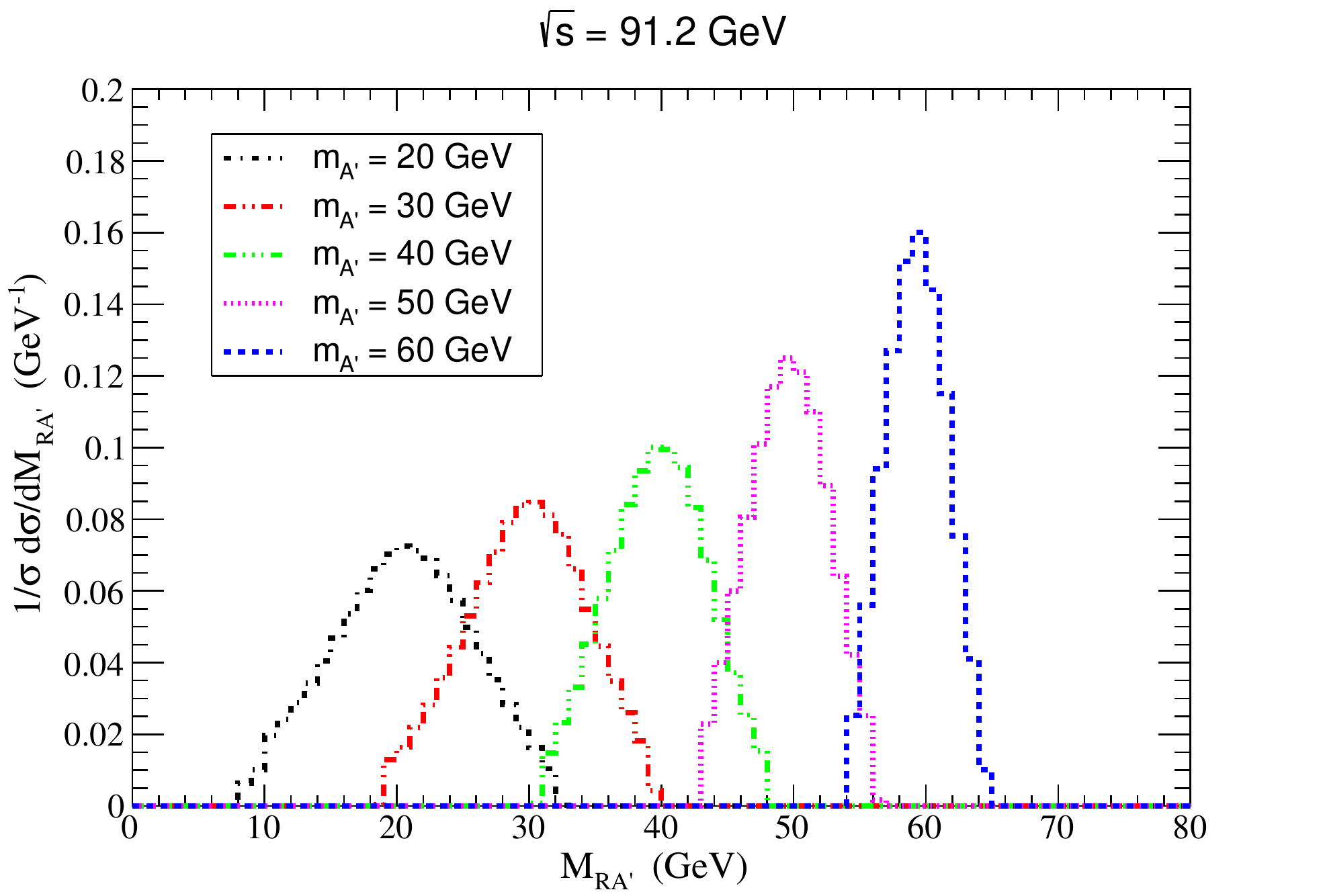}
\includegraphics[width=3.3in]{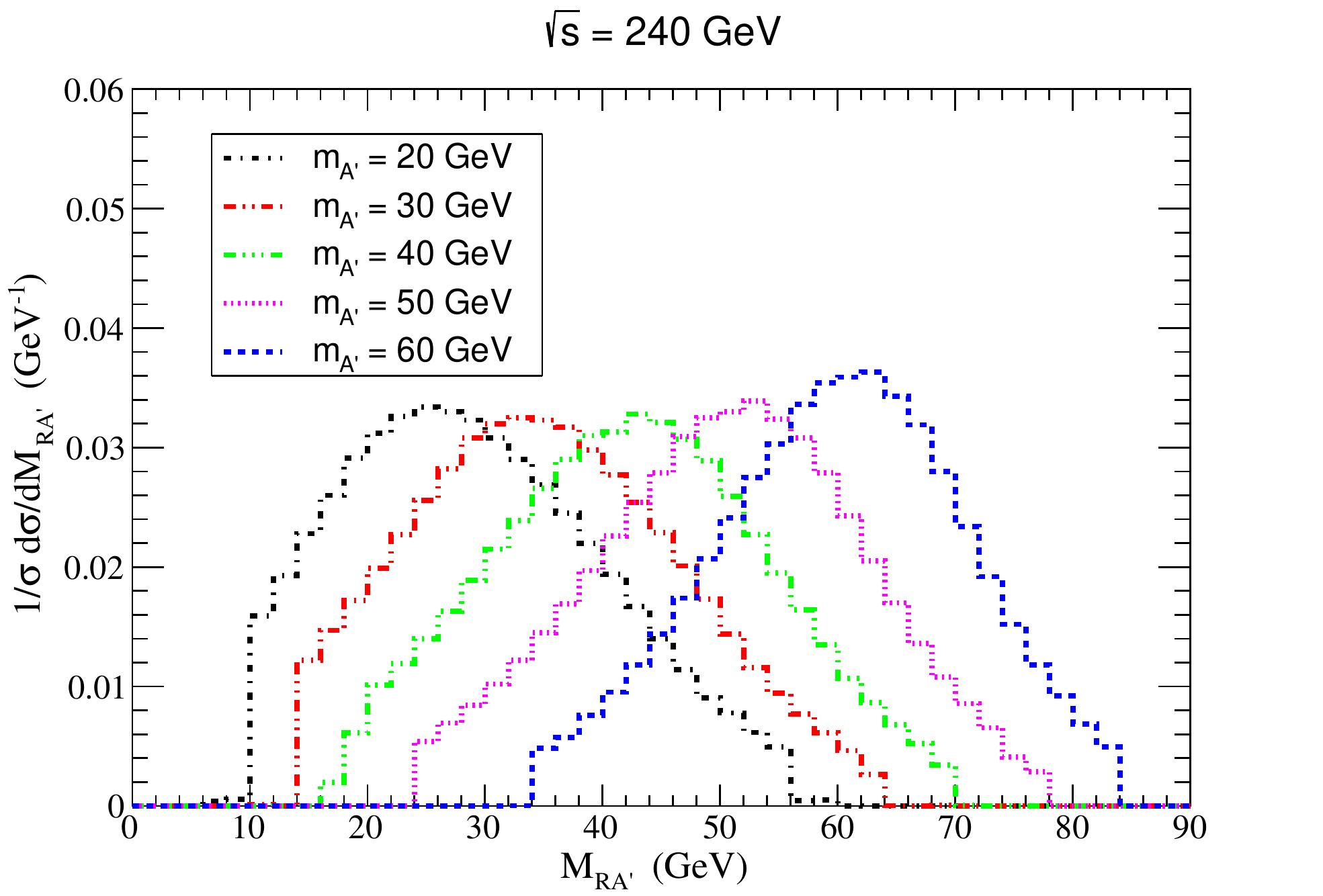}
\caption{Normalized differential cross-section $d\sigma/dM_{RA^{\prime}}$ as function of the invariant mass of the dark photon for $m_{A^{\prime}}=$ 20, 30, 40, 50 and 60~GeV, of the process $e^{+} e^{-}\rightarrow q \bar{q} A^{\prime}$ for $\sqrt{s}=$ 91.2~GeV (left panel) and $\sqrt{s}=$ 240~GeV (right panel), with the smearing and proper kinematic cuts.}
\label{fig:invariant mass_distribution} 
\end{figure}

In Fig.~\ref{fig:invariant mass_distribution}, we show the differential cross-section $d\sigma/dM_{RA^{\prime}}$ as function of the invariant mass of the dark photon for $m_{A^{\prime}}=$ 20, 30, 40, 50 and 60~GeV, with the smearing and the above kinematic cuts.
The reconstructed signal has a shape that complies with a Gaussian distribution with the expectation of $m_{A^{\prime}}$ and the standard deviation of the energy resolution of $\delta (E)$.
In contrast to the case of $\sqrt{s}=$ 91.2 GeV, the signal at $\sqrt{s}=$ 240 GeV has a wider spread since $\delta (E)$ is larger.

In order to identify the dark photon signal against the background, the significance of the signal-to-noise ratio needs to be explored. To enhance the significance, we impose the following cuts on the invariant mass spectrum: $|M_{RA^{\prime}} - m_{A^{\prime}}| <$ 6 GeV at $\sqrt{s} = $91.2 GeV, and $|M_{RA^{\prime}} - m_{A^{\prime}}| <$ 12 GeV at $\sqrt{s} = $240 GeV. For $\sqrt{s}=$ 91.2 GeV and with the CEPC integrated luminosity of $\cal L$ $=2 ~\rm{ab^{-1}}$ and for several $m_{A^{\prime}}$ values, we estimate the number of events for the signal ($N_S$) and background ($N_B$) processes, as well as the significance $ S/\sqrt{B}$, as listed in Table \ref{significance_91.2_table}. It can be seen that for $m_{A^{\prime}}=$ 20, 30, 40 and 50~GeV, the significance is greater than 3$\sigma$.

\begin{table}[htbp]
\caption{Number of events for the signal ($N_S$) and background ($N_B$) processes and the significance $S/\sqrt{B}$ for the integrated luminosity $\cal L$ $=2 ~\rm{ab^{-1}}$ at $\sqrt{s} = $91.2 GeV, with the smearing and proper kinematic cuts.}
\centering
\newcolumntype{d}{D{.}{.}{2}}
\begin{tabular}{cccccc}
  \hline
  $m_{A^{\prime}}$  &20 GeV& 30 GeV & 40 GeV & 50 GeV & 60 GeV  \\ \hline
  $N_{S}~(\cal L =\rm{2~ab^{-1}})$ & 191 & 368& 372 & 206& 46  \\
  $N_{B}~(\cal L =\rm{2~ab^{-1}})$ & 2503 & 3697 & 3636 & 2304& 799  \\
  $ S/\sqrt{B} $ & 3.82 & 6.05 &6.17 &4.29 & 1.63  \\ \hline
\end{tabular}
\label{significance_91.2_table}
\end{table}

\begin{table}[htbp]
\caption{The same as Table \ref{significance_91.2_table}, but for $\cal L$ $=20 ~\rm{ab^{-1}}$, $\sqrt{s} = $240 GeV and $|M_{RA^{\prime}} - m_{A^{\prime}}| <$ 12 GeV.}
\centering
\newcolumntype{d}{D{.}{.}{2}}
\begin{tabular}{cccccc}
  \hline
  $m_{A^{\prime}}$  &20 GeV& 30 GeV & 40 GeV & 50 GeV & 60 GeV  \\ \hline
  $N_{S}~(\cal L =\rm{20~ab^{-1}})$ & 2 & 10 & 23 & 39 & 53  \\
  $N_{B}~(\cal L =\rm{20~ab^{-1}})$ & 60252 &114953 & 210674 & 380295 & 682870 \\
  $ S/\sqrt{B} $ &0.00815 & 0.0295 & 0.0501 & 0.0632& 0.0641 \\ \hline
\end{tabular}
\label{significance_240_table}
\end{table}

\begin{figure}[htbp]
\centering
\includegraphics[width=3.5in]{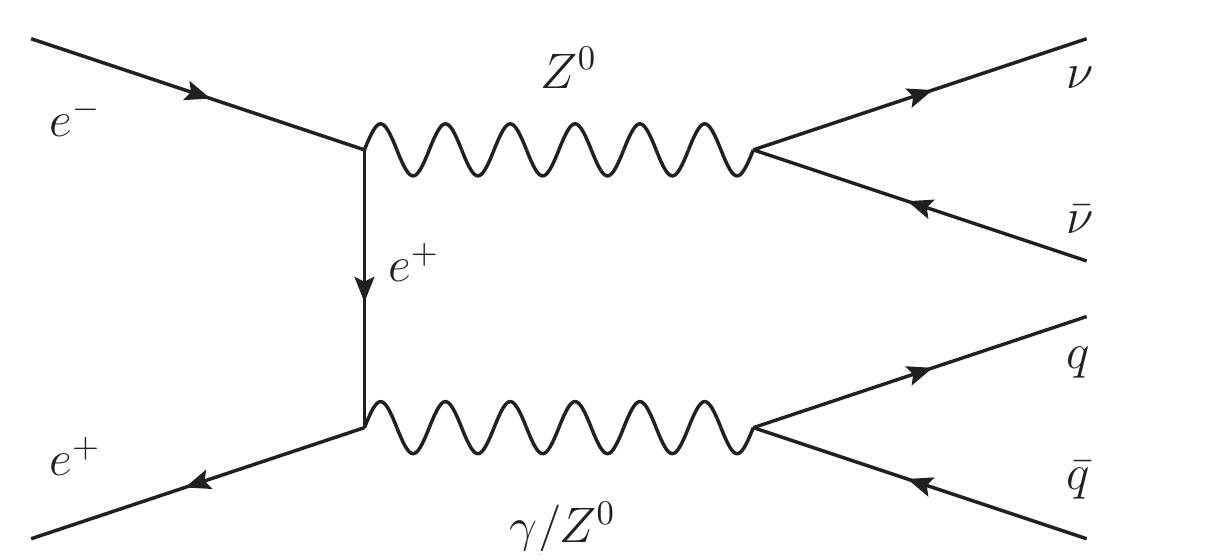}
\caption{A possible topology of the Feynman diagram for the background, which is excluded for the signal process.}
\label{fig:background_001} 
\end{figure}

In the case of the CEPC operating energy of $\sqrt{s}=$ 240 GeV, we adopt a higher integrated luminosity of $\cal L$ $=20 ~\rm{ab^{-1}}$. The number of events for the signal and background processes and the significance $ S/\sqrt{B}$ are given in Table \ref{significance_240_table}.
In comparison with Table \ref{significance_91.2_table}, we obtain a much smaller number of dark photon events. This is understandable since for $ 20~\rm{GeV} < m_{A^{\prime}} < 60~\rm{GeV}$, the cross-section decreases with the center-of-mass energy for $\sqrt{s}>$ 91.2 GeV, as demonstrated in Fig. \ref{fig:cross_section_v_s} (a) and (c). In addition, we obtain many more background events for $\sqrt{s}=$ 240 GeV than for $\sqrt{s}=$ 91.2 GeV. This is due to the new topology of Feynman diagram for the background process shown in Fig.~\ref{fig:background_001}, whose contribution increases with $\sqrt{s}$. This topology is excluded in the signal since we assumed that the dark photon interacts only with quarks.

\begin{figure}[htbp]
\centering
\includegraphics[width=3.4in]{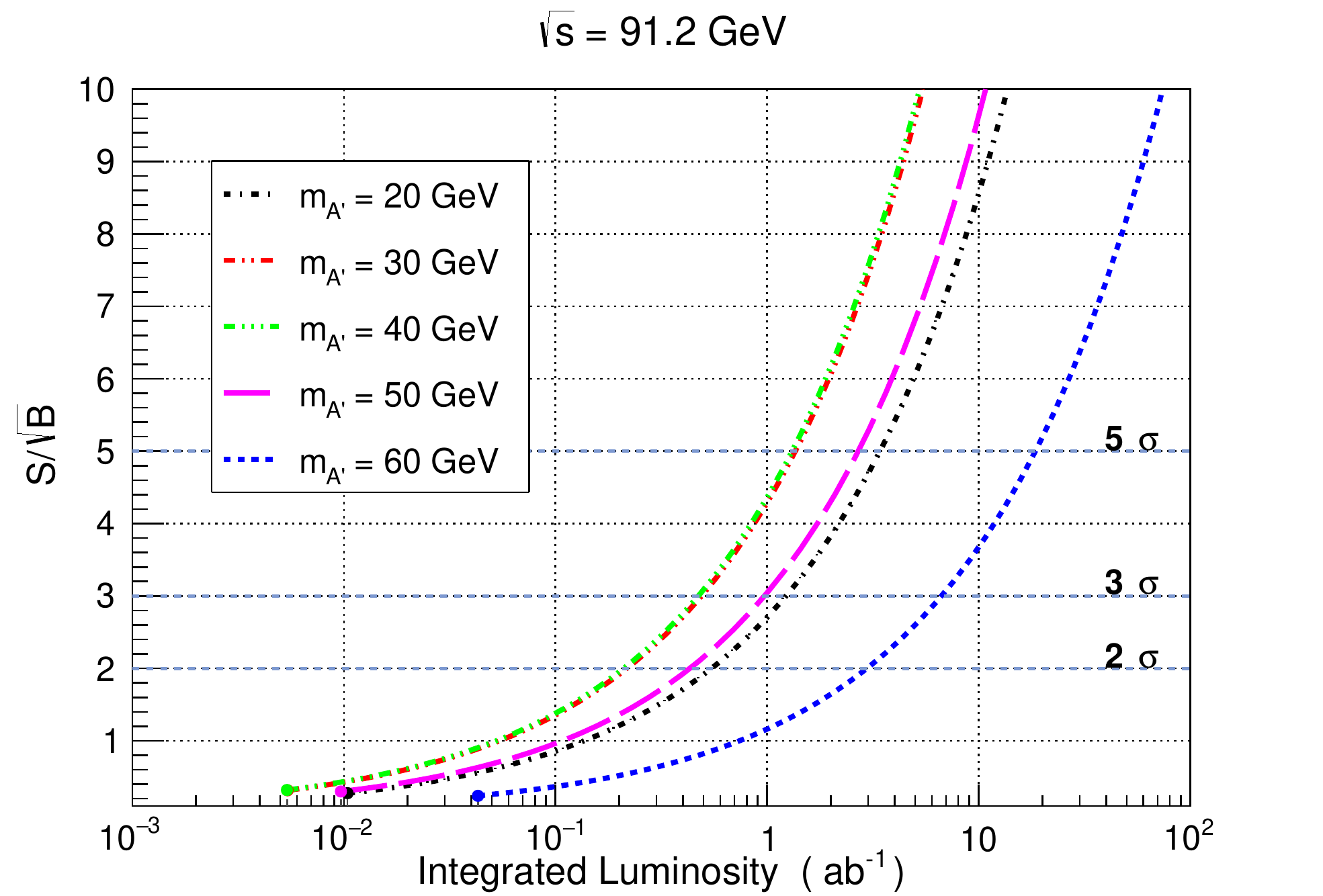}
\includegraphics[width=3.4in]{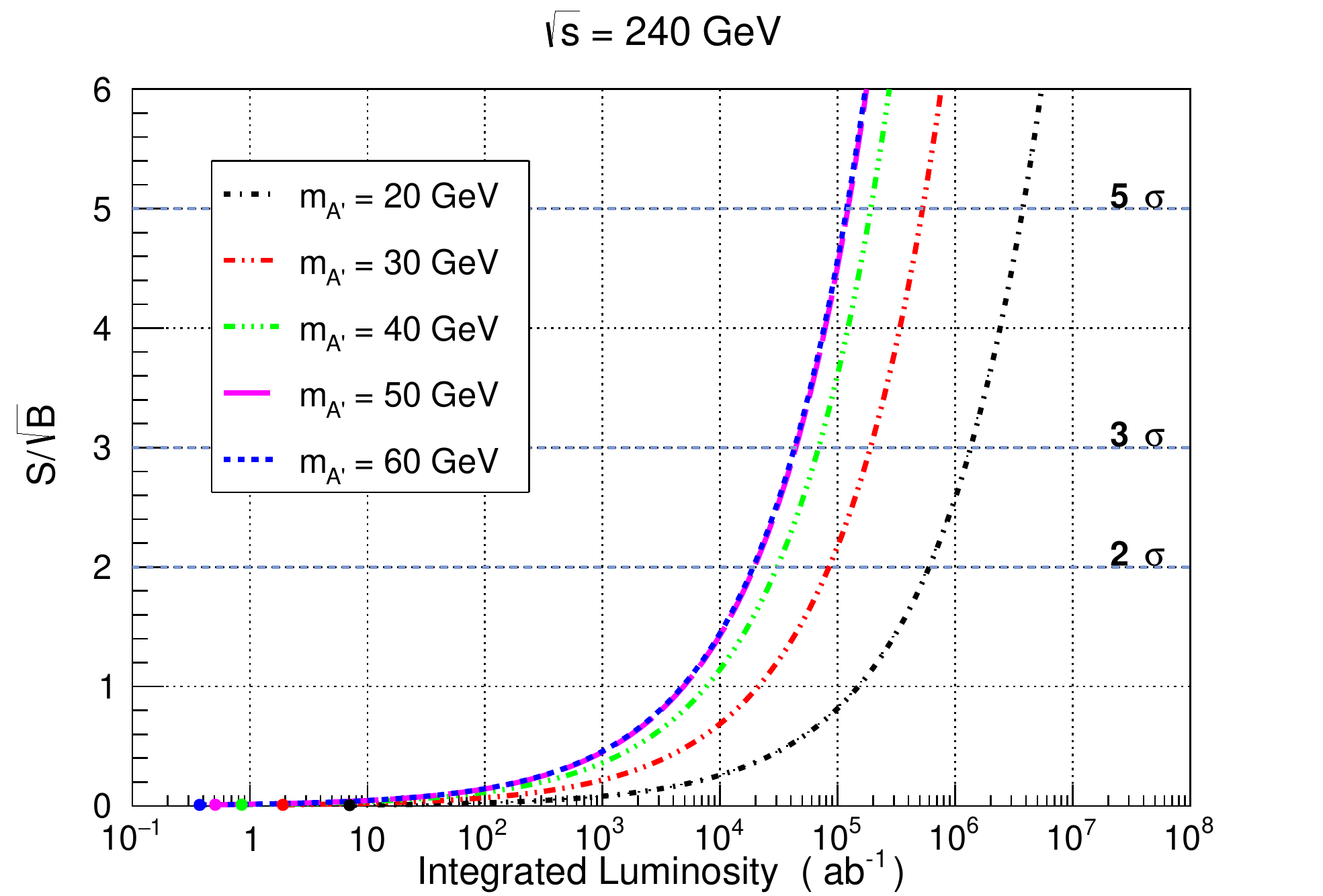}
\caption{Significance versus integrated luminosity for $m_{A^{\prime}}=$ 20, 30, 40, 50 and 60~GeV, and $\sqrt{s}$ = 91.2~GeV (left panel) and $\sqrt{s}$ = 240~GeV (right panel). The dots represent the minimum integrated luminosity for one signal event.}
\label{fig:significance_luminosity} 
\end{figure}

 As an additional element relevant for a future CEPC experiment, we present the significance $S/\sqrt{B}$ versus the integrated luminosity for $\sqrt{s} = $91.2 GeV and $\sqrt{s} = $240 GeV in Fig.~\ref{fig:significance_luminosity}.
In the case of $\sqrt{s} = $91.2 GeV, the minimum integrated luminosities for the 3$\sigma$ discovery of the dark photon with $m_{A^{\prime}}=$ 20, 30, 40, 50 and 60~GeV are 1.23, 0.490, 0.473, 0.971 and 6.67~$\rm{ab^{-1}}$, respectively.
Hence, it is understandable why the dark photon signal was not found at the Large Electron-Positron (LEP) collider, since the total luminosity of the LEP experiments~\cite{ALEPH:2005ab} did not reach the minimum integrated luminosity for the 3$\sigma$ discovery of the dark photon with 20~GeV$< m_{A^{\prime}}<$ 60~GeV.
At CEPC with $\sqrt{s} = $91.2 GeV, the yearly luminosity is expected to be $4~\rm{ab^{-1}year^{-1}}$ for a single interaction point (CEPC will have two interaction points), and it would be possible for a CEPC experiment to perform a decisive measurement of the  dark photon (20~GeV$< m_{A^{\prime}}<$ 60~GeV) in less than one operating year.
In the case of $\sqrt{s} = $ 240 GeV, the minimum integrated luminosities  required for one signal event with the above $m_{A^{\prime}}$ values are 7.06, 1.91, 0.853, 0.508 and 0.374~$\rm{ab^{-1}}$, respectively. Therefore, with CEPC running at $\sqrt{s} = $ 240 GeV and a luminosity of $0.4~\rm{ab^{-1}year^{-1}}$ for a single interaction point, it would be hardly possible to get any signal of the dark photon (20~GeV$< m_{A^{\prime}}<$ 60~GeV) in one operating year.

\section{Summary}

The dark sector may consist of not only DM but also of one or more new force-carrying mediators which couple to the SM particles.  We discussed the vector dark photon $A^{\prime}$ and the scalar mediator $\phi$ which could be produced in the processes $e^{+} e^{-} \rightarrow q \bar{q} A^{\prime}$ and $e^{+} e^{-} \rightarrow q \bar{q} \phi$ at future $e^{+} e^{-}$ colliders. The production cross-sections of these processes were predicted for $\sqrt{s}$ = 91.2~GeV, 240~GeV, 500~GeV and 1~TeV. We further studied the kinematic distributions of the two-jet system in the final state, and found that they could be used to identify (or exclude) the dark photon and the dark scalar mediator, as well as to distinguish between them.
In this work, we only considered the interaction between the dark photon and quarks, and with the process $e^{+} e^{-} \rightarrow q \bar{q} A^{\prime}$ as an example, we investigated the discovery potential of the dark photon at CEPC with $\sqrt{s}$ = 91.2 GeV and 240 GeV. It was shown that the dark photon with $m_{A^{\prime}}$ ranging from 20 GeV to 60 GeV might be discovered in the process $e^{+} e^{-} \rightarrow q \bar{q} A^{\prime}$ at $e^+e^-$ colliders, e.g. at the super-Z factory or CEPC, with the minimum required integrated luminosity for the 3$\sigma$ discovery of about 0.473~$\sim$ 6.67~$\rm{ab^{-1}}$.
If the interaction between the dark mediator and leptons is also considered, $e^{+} e^{-} \rightarrow \ell^{+} \ell^{-} A^{\prime}$ and $e^{+} e^{-} \rightarrow \gamma A^{\prime}$ could be the other interesting processes to study, where the $WW$ production would be the background. The method proposed in this work could also be used to search for any other invisible particles in $e^{+} e^{-}$ annihilation.

\section*{Acknowledgements}

This work was supported in part by the National Natural Science Foundation of China (grant Nos. 11875179, 11325525, 11635009, 11775130 and 11905112), the Natural Science Foundation of Shandong Province (grant Nos. ZR2017MA002, ZR2019QA012) and the Fundamental Research Funds of Shandong University (grant No. 2019GN038).

\end{document}